\documentclass[sn-nature]{sn-jnl}

\usepackage[utf8]{inputenc}
\DeclareUnicodeCharacter{2212}{-}
\usepackage[textsize=small]{todonotes}
\usepackage{graphicx}%
\usepackage{multirow}%
\usepackage{amsmath,amssymb,amsfonts}%
\usepackage{amsthm}%
\usepackage{mathrsfs}%
\usepackage[title]{appendix}%
\usepackage{xcolor}%
\usepackage{textcomp}%
\usepackage{manyfoot}%
\usepackage{booktabs}%
\usepackage{algorithm}%
\usepackage{algorithmicx}%
\usepackage{algpseudocode}%
\usepackage{listings}%
\usepackage[printonlyused,nohyperlinks]{acronym}%

\begin{document}

\title[Accelerating materials discovery with AI and Cloud HPC]{Accelerating computational materials discovery with artificial intelligence and cloud high-performance computing: from large-scale screening to experimental validation}


\author[1]{\fnm{Chi} \sur{Chen}}
\equalcont{These authors contributed equally to this work.}

\author[2]{\fnm{Dan Thien} \sur{Nguyen}}
\equalcont{These authors contributed equally to this work.}

\author[2]{\fnm{Shannon J.} \sur{Lee}}

\author[1]{\fnm{Nathan A.} \sur{Baker}}

\author[2]{\fnm{Ajay S.} \sur{Karakoti}}

\author[1]{\fnm{Linda} \sur{Lauw}}

\author[3]{\fnm{Craig} \sur{Owen}}

\author[2]{\fnm{Karl T.} \sur{Mueller}}

\author[1]{\fnm{Brian A.} \sur{Bilodeau}}

\author*[2]{\fnm{Vijayakumar} \sur{Murugesan}}\email{vijay@pnnl.gov}

\author*[1]{\fnm{Matthias} \sur{Troyer}}\email{matthias.troyer@microsoft.com}

\affil[1]{\orgdiv{Azure Quantum}, \orgname{Microsoft}, \orgaddress{\street{One Microsoft Way}, \city{Redmond}, \state{WA}, \postcode{98052}, \country{USA}}}

\affil[2]{\orgdiv{Physical and Computational Sciences Directorate}, \orgname{Pacific Northwest National Laboratory}, \orgaddress{\street{902 Battelle Blvd.}, \city{Richland}, \state{WA}, \postcode{99352}, \country{USA}}}

\affil[3]{\orgdiv{Microsoft Surface}, \orgname{Microsoft}, \orgaddress{\street{One Microsoft Way}, \city{Redmond}, \state{WA}, \postcode{98052}, \country{USA}}}


\newcommand{\naseries}{Na$_x$Li$_{3-x}$YCl$_6$}


\abstract{
High-throughput computational materials discovery has promised significant acceleration of the design and discovery of new materials for many years.
Despite a surge in interest and activity, the constraints imposed by large-scale computational resources present a significant bottleneck.
Furthermore, examples of large-scale computational discovery carried through experimental validation remain scarce, especially for materials with product applicability.
Here we demonstrate how this vision became reality by first combining state-of-the-art \ac{AI} models and traditional physics-based models on cloud \ac{HPC} resources to quickly navigate through more than 32 million candidates and predict around half a million potentially stable materials.
By focusing on solid-state electrolytes for battery applications, our discovery pipeline further identified 18 promising candidates with new compositions, and rediscovered a decade's worth of collective knowledge in the field as a byproduct. By employing around one thousand virtual machines (VMs) in the cloud, this process took less than 80 hours.
We then synthesized and experimentally characterized the structures and conductivities of our top candidates, the \naseries\ (0 $< x <$ 3) series, demonstrating the potential of these compounds to serve as solid electrolytes. Additional candidate materials that are currently under experimental investigation could offer more examples of the computational discovery of new phases of Li- and Na-conducting solid electrolytes. We believe that this unprecedented approach of synergistically integrating \ac{AI} models and cloud \ac{HPC} not only accelerates materials discovery but also showcases the potency of \ac{AI}-guided experimentation in unlocking transformative scientific breakthroughs with real-world applications.
}




\maketitle


\section{Introduction} \label{sec:intro}

Materials discovery has been integral to human progress throughout history, marking pivotal shifts in civilization from the stone age to the modern era.
Traditionally, materials have been discovered predominantly through exploration and trial-and-error laboratory experiments.
In the last few decades, computational methods have become increasingly popular for predicting and understanding material properties. 
However, the goal of discovering materials with targeted properties \textit{in silico} has yet to be realized. 
In recent years, advancements in computational science and \ac{AI}, together with curated datasets emerging in public databases like the Materials Project~\cite{jain_commentary_2013}, AFLOWLIB~\cite{curtarolo_aflowliborg_2012}, Open Quantum Materials Database~\cite{saal_materials_2013}, Materials Cloud~\cite{talirz_materials_2020}, and others have created opportunities for a new age of materials discovery, where computational approaches take the lead in the prediction-synthesis-characterization cycle. 

\Ac{ML} models for materials science have the potential to vastly expedite the computational discovery process.
State-of-the-art \ac{ML} models can predict the results of physics-based quantum mechanical calculations but are several orders of magnitude faster, making them ideal for predicting general material properties~\cite{chen_graph_2019,xie_crystal_2018,choudhary_atomistic_2021}. In addition to direct property prediction, combining universal \ac{ML} potentials such as M3GNet~\cite{chen_universal_2022}, CHGNet~\cite{deng_chgnet_2023}, and GNoME~\cite{merchant_scaling_2023} has made it possible to perform geometric optimization, and hence evaluate thermodynamic stability, for arbitrary combinations of elements and structures. 
The significant speed advantage of \ac{ML}-based techniques over direct simulation has made it possible to explore materials across a vast chemical space that greatly exceeds the number of known materials.

Some recent studies have produced massive amounts of data on hypothetical materials~\cite{chen_universal_2022,merchant_scaling_2023}, but without experimental validation.
Autonomous labs have shown early success in synthesizing materials~\cite{szymanski_autonomous_2023}, but the technical validation of those materials' functional properties remains to be demonstrated.
Discovering technologically relevant materials is likely orders of magnitude harder than discovering synthesizable materials because the former are usually outliers in the materials distribution with unique, sometimes counter-intuitive, combinations of properties~\cite{zunger_inverse_2018}.

Even with the acceleration offered by \ac{ML}-based potentials over quantum mechanical energy and force calculations, the lack of required large-scale computational resources is a challenge that has slowed progress.
While it is relatively trivial to generate $10^9$ material candidates, the evaluation of those candidates still requires significant computing resources.
This challenge can be overcome through cloud \ac{HPC}, the scale of which has recently become widely known due to its ability to train and host large-scale \ac{AI} models like GPT-4, which require a massive number of \acp{GPU}.
In the realm of computational materials discovery, these advancements imply that cloud computing can effectively manage smaller-scale, complex calculation jobs, greatly increasing the number of material candidates that can be evaluated computationally.
Additionally, these reproducible cloud-based workflows substantially reduce entry barriers for new users, thereby democratizing the materials discovery process.

In this work, we demonstrate the utility and impact of the combined strengths of \ac{AI} and cloud \ac{HPC} to design and discover materials.
We showcase the discovery and experimental validation of new material compositions with targeted properties by filtering a vast pool of potential solid-state electrolytes for battery technology applications.
The outcomes of this process are noteworthy for their rapid identification of promising material candidates, reflecting insights that have emerged over the last decade in this field.
This efficient screening and discovery process significantly reduces the overall time from concept to solution in materials discovery.

\section{Large-scale materials discovery}

\subsection{Exploring a vast chemical space}

\begin{figure}
    \centering
    \includegraphics[width=0.8\textwidth]{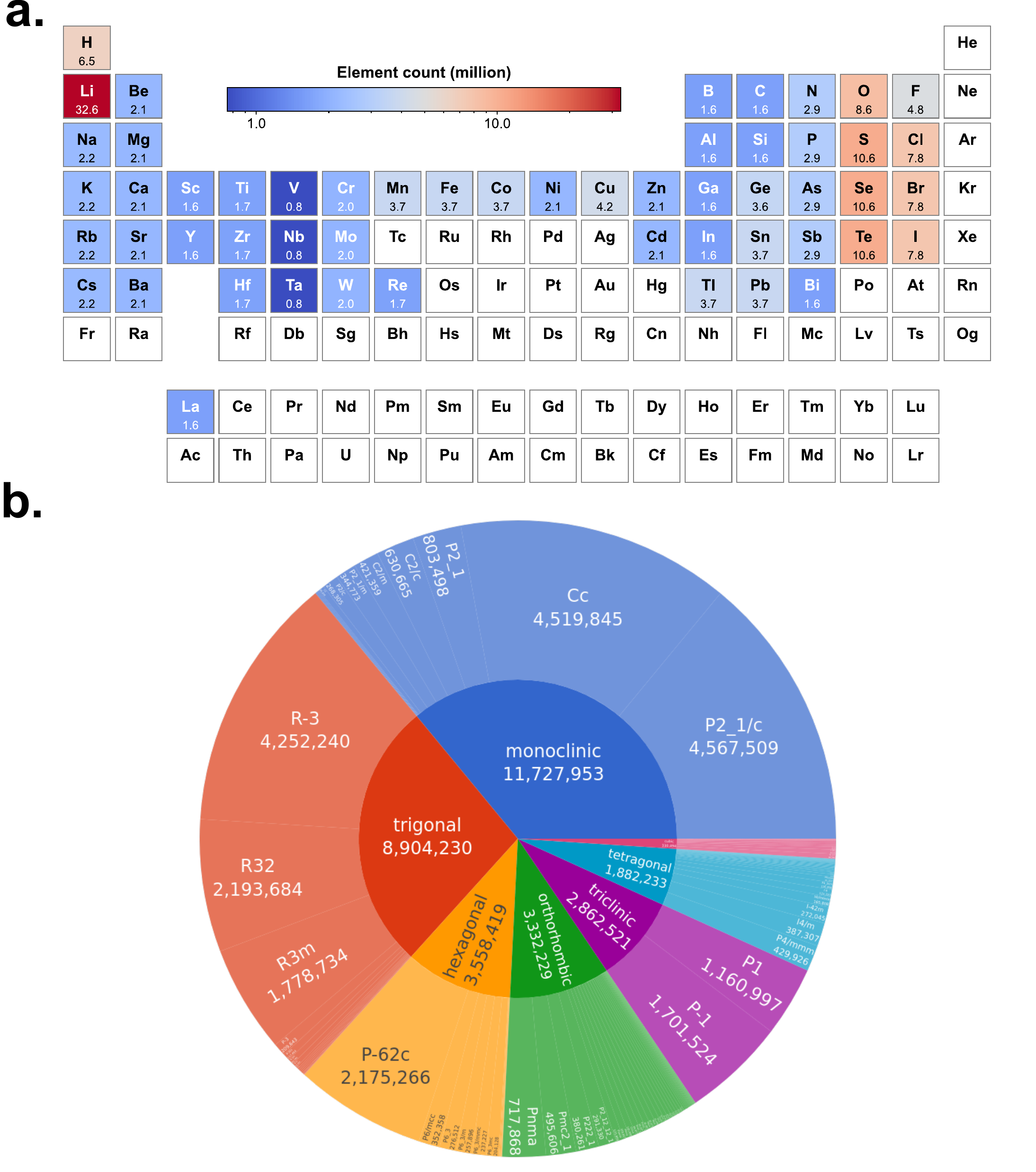}
    \caption{Distribution of substituted material composition and structure.
    a.\ Element distribution and b.\ space group distribution for the initial candidates.
    Visualizations were created with pymatviz~\cite{riebesell_pymatviz_2022}.} \label{fig:chemical-space}
\end{figure}

Structure candidates were generated based on ionic substitution to known crystal structures.
This approach has previously demonstrated success in helping discover new materials in various applications~\cite{hautier_data_2011, wu_first_2013, wang_mining_2018, zhu_li3yps42_2017}.
However, unlike previous approaches, the current work does not constrain substitution by ionic similarity learned from mining materials databases~\cite{hautier_data_2011} and therefore has a broader search space.
We chose 54 elements, indicated by the colored cells in Figure~\ref{fig:chemical-space}a, as our starting point for the most likely compositions of Li-conducting solid-state electrolytes.
We chose to remove later-row transition metals and most lanthanides because those are not commonly found in Li battery materials.
Noble gases and actinides are typically not part of viable functional materials applications and hence were not included.

Common oxidation states of the 54 elements were obtained from Pymatgen~\cite{ong_python_2013} data.
To generate the initial structure candidates, we used the 2022 version of the \ac{ICSD}~\cite{belsky_new_2002}, and then used Pymatgen~\cite{ong_python_2013} StructureMatcher to match similar structures and obtain structure prototypes.
We used only the ordered structures with integer site occupancies that quantum mechanical calculations can handle, and performed iso-valent substitutions to those prototypes, which generated 32,598,079 initial candidates.
The space groups of those candidates are shown in Figure~\ref{fig:chemical-space}b.
Our candidate materials cover 184 out of 230 space groups and represent all seven crystal systems.
Compared to the candidates generated in matterverse.ai~\cite{chen_universal_2022}, the current approach expands the number of prototype structures, does not have a 50-atom limit, and is not limited to a specific group of anions.

The distribution of candidates in this study covers most of the technically interesting elements and space groups. 
Notably, the substitution approach used here to identify candidate structures is intentionally broad but is not necessarily thorough for a given composition.
This contrasts with other prediction approaches for crystal structures such as AIRSS~\cite{pickard_high-pressure_2006}, USPEX~\cite{glass_uspexevolutionary_2006}, and CALYPSO~\cite{wang_calypso_2012}, which have a narrower scope but are more thorough.
Our broader search strategy has a less-biased focus on specific regions of material space and therefore should lead to a more diverse selection of new materials.

\subsection{Screening workflow for stable materials and solid electrolytes}

\begin{figure}
    \centering
    \includegraphics[width=\textwidth]{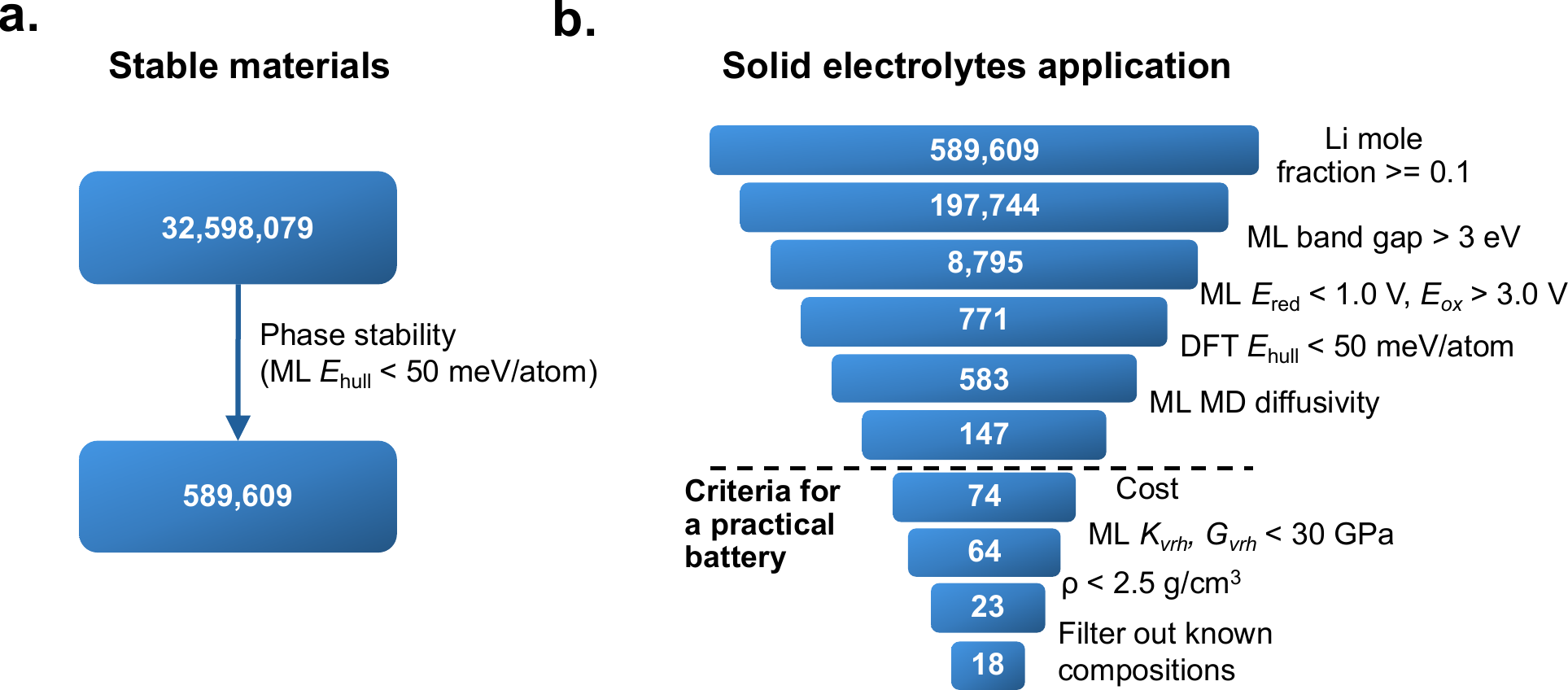}
    \caption{The main screening stages of the workflow for filtering candidate materials.
    Of the final 23 candidates, 18 materials have compositions that have not been previously reported.} \label{fig:screening-workflows}
\end{figure}

This large space of materials candidates was first screened for stability, producing 589,609 candidates that were predicted to be stable. From this pool of candidates, we selected materials with desired properties for solid-state electrolytes. Our funnel-based screening workflow started with property filters that use \ac{AI} models to quickly screen initial candidates, and then applied physics-based \ac{DFT} calculations to serve as high-accuracy filters on the most promising candidates (Figure~\ref{fig:screening-workflows}).
In general, a filtering process should start with the fastest calculations to focus computational effort on the top candidates.

Figure~\ref{fig:screening-workflows} illustrates these two workflow stages.
The first stage, shown in Figure~\ref{fig:screening-workflows}a, identified stable materials using \ac{ML} potentials to perform structural relaxation and assess thermodynamic phase stabilities using pymatgen~\cite{ong_python_2013} and Materials Project reference data~\cite{jain_commentary_2013}.
For compositions with multiple polymorphic structures, we kept only the lowest energy structure.
This step reduced the number of candidates from 32.6 million to 589,609 materials that were predicted to be stable.

The second stage of the workflow, shown in Figure~\ref{fig:screening-workflows}b, filtered this large pool of materials to discover those with electronic and electrochemical properties that are suitable for solid electrolytes in battery applications.
We first filtered the candidates to include materials with Li site mole fractions of at least 0.1 to ensure a reasonable number of sites that support Li$^+$ conduction.
Then we used the band gap to screen for electronic insulators and provide an upper limit for the \ac{ESW}; this filtering retained candidates that had an \ac{ML}-predicted band gap of greater than 3 eV.
Next, we applied a more accurate \ac{ESW} filter by calculating the redox potential from the \ac{ML}-predicted energies and retained only those candidates with energetically unfavorable reduction ($E_{\text{red}} < 1$ V w.r.t.~Li/Li$^+$) and oxidation ($E_{\text{ox}} > 3$ V w.r.t.~Li/Li$^+$).

After the \ac{AI}-based step, we used \ac{HPC} simulations,  starting with \ac{DFT} relaxation and static calculations to confirm the stability predicted by \ac{ML}, and performed \ac{MD} simulations using \ac{ML} potentials.
We assessed Li conductivity using Li diffusivity as a proxy, following the approach by Zhu et al.~\cite{zhu_li3yps42_2017}, where Li diffusivity was calculated at 800 K and 1200 K.
We applied a diffusion constant filter of $D(1200\text{K})/D(800\text{K}) < 7$, which corresponds roughly to a Li migration barrier of less than 0.4 eV, and $D(800\text{K}) > 10^{-5}$ cm$^2$ s$^{-1}$.
We ended up with 147 promising candidates that passed all criteria. 

In the latter half of the second stage (Figure~\ref{fig:screening-workflows}b), we filtered the candidates to select only those with properties needed to integrate each material into \emph{practical} batteries.
We aimed to reduce the cost by removing Be, Sc, Cs, Rb, and Hf; this filter also served as a proxy for filtering by mineral abundance.
By predicting and removing candidates with high bulk and shear moduli (greater than 30 GPa), we retained materials that were soft and could potentially form a cohesive interface with cathodes and anodes.
Finally, we removed high density materials (greater than 2.5 g cm$^{-3}$) because mass density is inversely related to energy density.
Together, the two workflow stages identified 23 final candidates as well as a large dataset of properties for the original pool of 32,598,079 candidates.
Eighteen of the 23 final candidates have compositions that have not been previously reported.
To account for possible disordered phases of the final candidates, we calculated structure energies for enumerated ordered structures of Na-Li compounds from partial substituted disordered structures, and kept the lowest energy structures.    

\subsection{Infrastructure for cloud-based materials discovery}

\begin{figure}
    \centering
    \includegraphics[width=\textwidth]{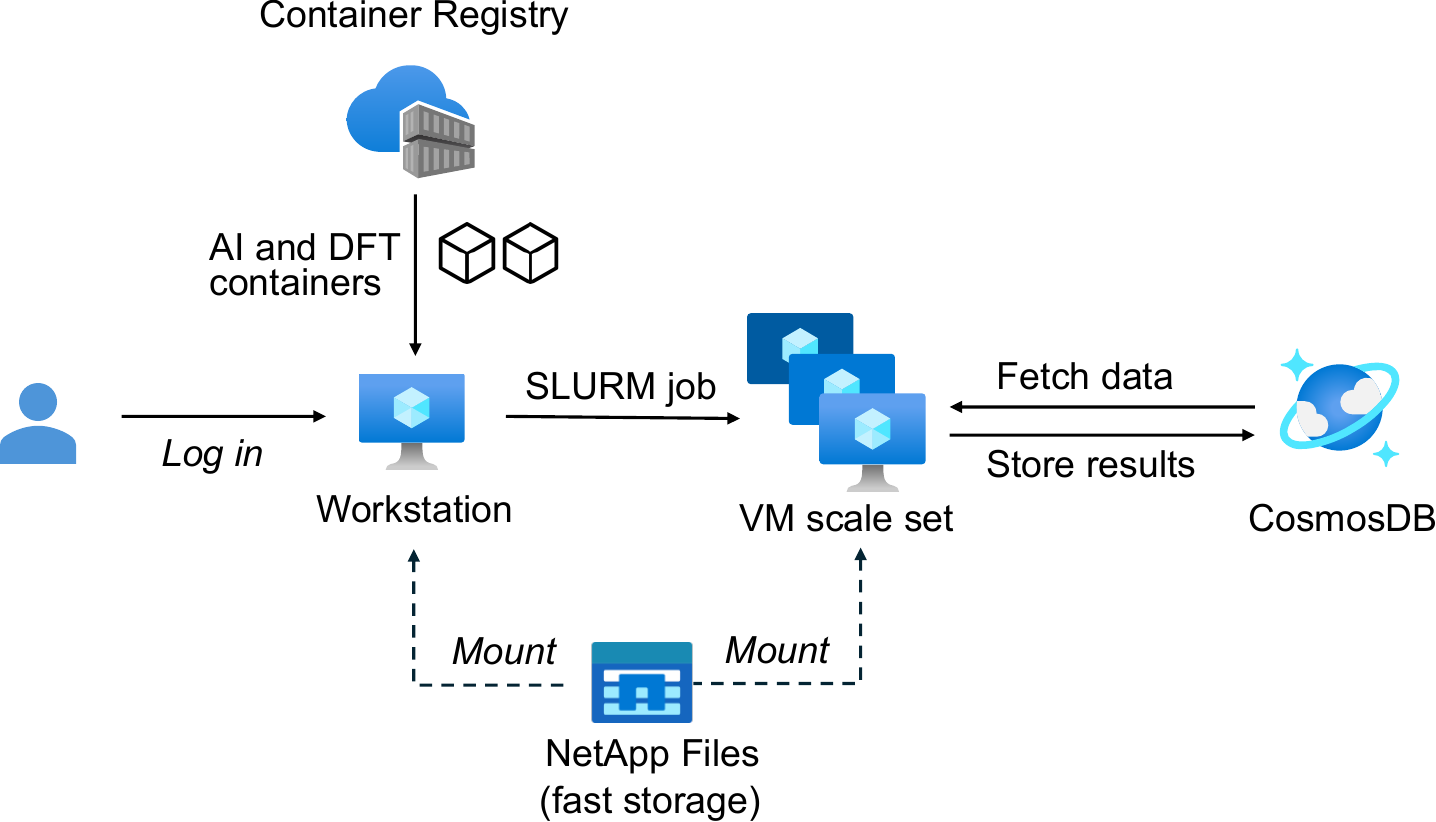}
    \caption{Schematic of the cloud environment used for materials discovery workloads in this study.} \label{fig:architecture}
\end{figure}

The Microsoft Azure Quantum team designed a cloud \ac{HPC} system based on the Azure \ac{HPC} On-Demand Platform \cite{azhop}.
The latest implementation of this infrastructure is available through {Azure Quantum Elements} \cite{aqe}.
We used easily accessible cloud-based resources of Azure Quantum Elements to perform all tasks from simple Python scripting to complex \ac{DFT} and \ac{MD} calculations.
Figure~\ref{fig:architecture} shows the core infrastructure of our large-scale computational materials discovery workflow.

The \ac{ML} models and \ac{DFT} code were built into Docker \cite{docker} container images and hosted in a private Azure Container Registry.
At runtime, a workstation \ac{VM} fetched the container images to NetApp Files \cite{netapp} storage mounted to the workstation and job queues or \ac{VM} scale sets.
We submitted computational jobs that run the containers to the \ac{VM} scale sets via the SLURM job scheduler \cite{slurm}.
The \ac{VM} scale set is the counterpart of a partition or a queue in conventional \ac{HPC}, and the number of \acp{VM} is adjustable at deployment time given sufficient quota.
Each job runs a container and is thus separate from the other jobs.

Data and metadata, such as job status and job chunk status, were kept in a searchable database hosted in CosmosDB \cite{cosmosdb} and accessed via a MongoDB API \cite{mongodb_api}.
Status tracking is essential for taking advantage of pre-emptible or low-priority \acp{VM} in the cloud, and for making sure the job can be restarted.
In this study, we used approximately one thousand \acp{VM} to run \ac{ML} potential-based optimization and \ac{DFT} jobs.

\subsection{Screening results and trends}

By using approximately 1,000 \acp{VM} for less than 80 hours, this filtering workflow created one of the largest datasets for \ac{ESW} of 8,795 materials and produced diffusivity data on a diverse set of 583 material compositions and structures.
This information will be important to guide the future design of electrolytes.

\begin{figure}
    \centering
    \includegraphics[width=0.95\textwidth]{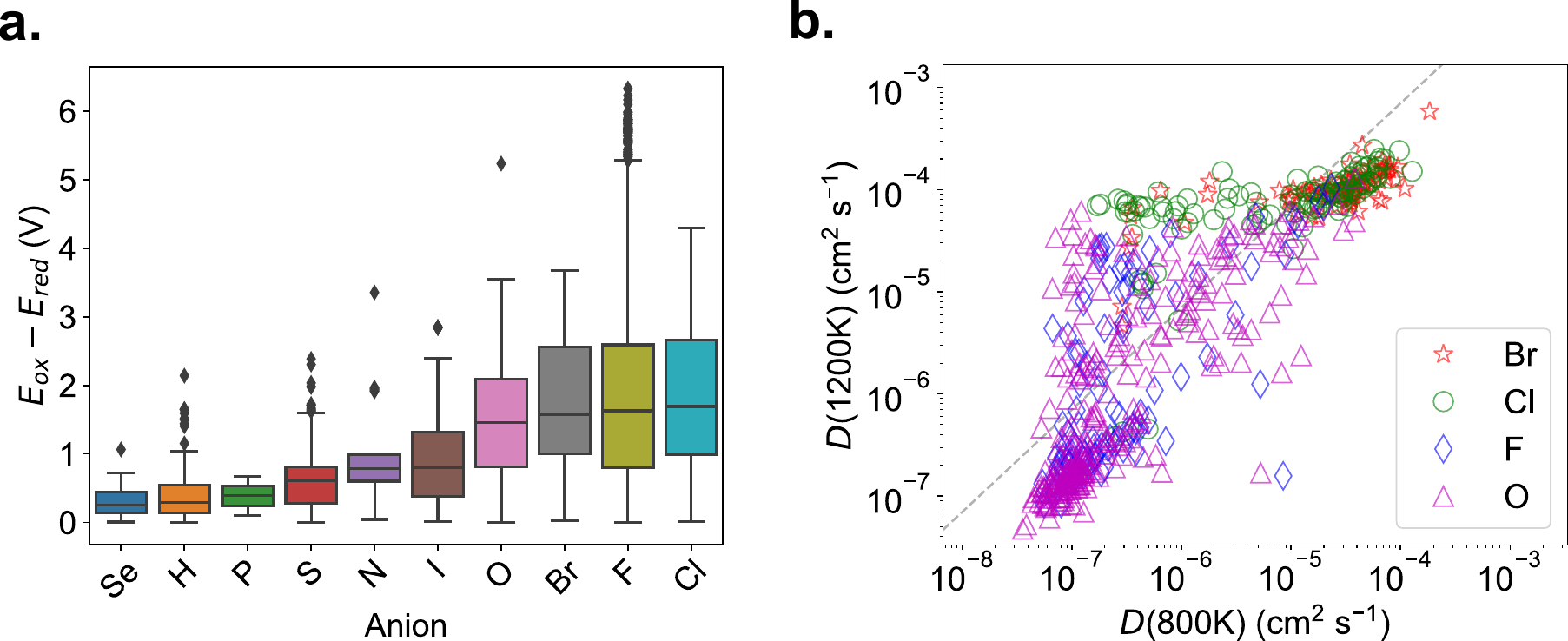}
    \caption{Candidate electrochemical and diffusivity properties.
    a.\ The \ac{ESW} ($E_{\text{ox}} - E_{\text{red}}$) for the wide band gap of 8,795 candidates.
    b.\ The self-diffusivity of Li in the materials at 800 K and 1200 K grouped by types of anions for 583 materials.
    The dashed line represents a ratio of $D(1200 \text{K})/D(800 \text{K})$ of 7, which corresponds to an activation energy of approximately 0.4 eV~\cite{zhu_li3yps42_2017}.
    Candidates below the dashed line have low activation energies and are thus preferred.} \label{fig:esw-diffusivity}
\end{figure}

Figure~\ref{fig:esw-diffusivity}a shows the \ac{ESW} for 8,795 wide band gap materials, grouped by the presence of specific anions.
For multi-anion materials, we selected the highest electronegativity anion as the representative one.
We found that early row halides such as F, Cl and Br tend to have  larger \ac{ESW}.
Oxides follow closely, but other chalcogenides such as S and Se have very small \acp{ESW}.
The trend matched well with a previous study from Wang {\it et al.}~\cite{wang_lithium_2019} on  Li$-M-X$ ternary systems, where $M$ = metal cation, $X$ = F, Cl, Br, I, O, S.
The \ac{ESW} trend suggests that chlorides, fluorides, and bromides are better candidates for solid electrolytes when electrochemical stability is a requirement. 

The 583 candidates that passed the \ac{ESW} filters all contained either Br, Cl, F, or O.
The Li diffusivities of the materials at 800 K and 1200 K are plotted in Figure~\ref{fig:esw-diffusivity}b and were used to select  materials with sufficiently low activation energies for Li movement across a range of temperatures.
Oxides and fluoride materials tend to have diffusivities that vary widely across compound species.
This variation is likely due to the sensitivity of Li conductivities to structural changes.
For example, even for the same composition, the tetragonal phase Li$_7$La$_3$Zr$_2$O$_{12}$ has much lower Li conductivity compared to the cubic phase~\cite{tan_synthesis_2011}.
Bromides and chlorides tend to have consistently good high-temperature diffusivities, and hence  are promising families of compounds to explore as electrolytes. 

\subsection{Final candidates}

Ultimately, we focused on four rare-earth metal-based halides for experimental validation (Table~\ref{tab:candidates}).
These halides included three candidates that passed through all property filters in our workflow, Li$_5$YCl$_8$, Li$_7$Y$_2$Cl$_{13}$, and Na$_2$LiYCl$_6$, as well as a Pna2$_1$ polymorph of known composition Li$_3$YCl$_6$, which passed the property-based filters.
The P$\bar{3}$m1 phase of Li$_3$YCl$_6$ is a well-known material with Li and Y site disorder~\cite{schlem_insights_2021,asano_solid_2018,hu_revealing_2023,steiner_neue_1992}.
We generated ordered enumerated structures from the known P$\bar{3}$m1 phases, but the energies of those structures were much higher.
Hence, we included the predicted Pna2$_1$ phase in our study to explore additional properties of this structure.

\begin{table}[b]
    \caption{Rare-earth metal-based halides selected for further investigation in this study.} \label{tab:candidates}
    \begin{tabular}{@{}llp{2cm}llll@{}}
        \toprule
        Composition & Space group & $E_{\text{hull}}$ (meV/atom) & Band gap (eV) & $E_\text{red}$ (V) & $E_\text{ox}$ (V) &$\sigma_{300 \text{K}}$ (mS/cm) \\
        \midrule
        Li$_3$YCl$_6$ & Pna2$_1$ & 24 & 4.91 & 0.65 & 4.27 & 12.17 \\
        Li$_5$YCl$_8$ & Cmmm & 32 & 4.99 & 0.65 & 4.27 & 1.12 \\
        Li$_7$Y$_2$Cl$_{13}$ & Pn$\bar{3}$ & 39 & 5.06 & 0.65 & 4.27 & 1.70 \\
        Na$_2$LiYCl$_6$ & R$\bar{3}$ & 15 & 5.20 & 0.65 & 3.80 & 0.12 \\
        \botrule
    \end{tabular}
\end{table}

All materials in Table~\ref{tab:candidates} have good \ac{DFT} \acp{ESW} (Figure~\ref{fig:top-esw}), large \ac{DFT} band gaps (Supplementary Figure~\ref{fig:top-band-gaps}), and good \ac{AIMD} Li conductivities at room temperature (Supplementary Figure~\ref{fig:top-diffusivities}).
Of particular interest is Na$_2$LiYCl$_6$ because of its Li-Na dual conductivity properties (see Supplementary Figure~\ref{fig:top-diffusivities} and Supplementary Figure~\ref{fig:n2116-na-diffusitivity}).
Both from a fundamental scientific perspective and in terms of technological applications, discovering a structural framework that enables the co-diffusion of two distinct alkali cations holds potential to greatly expand the design formulations of solid-state electrolyte materials.

\subsection{Synthesis and experimental characterization}

\subsubsection{Structure and composition}

\begin{figure}[h]
    \centering
    ~\includegraphics[width=1.0\textwidth]{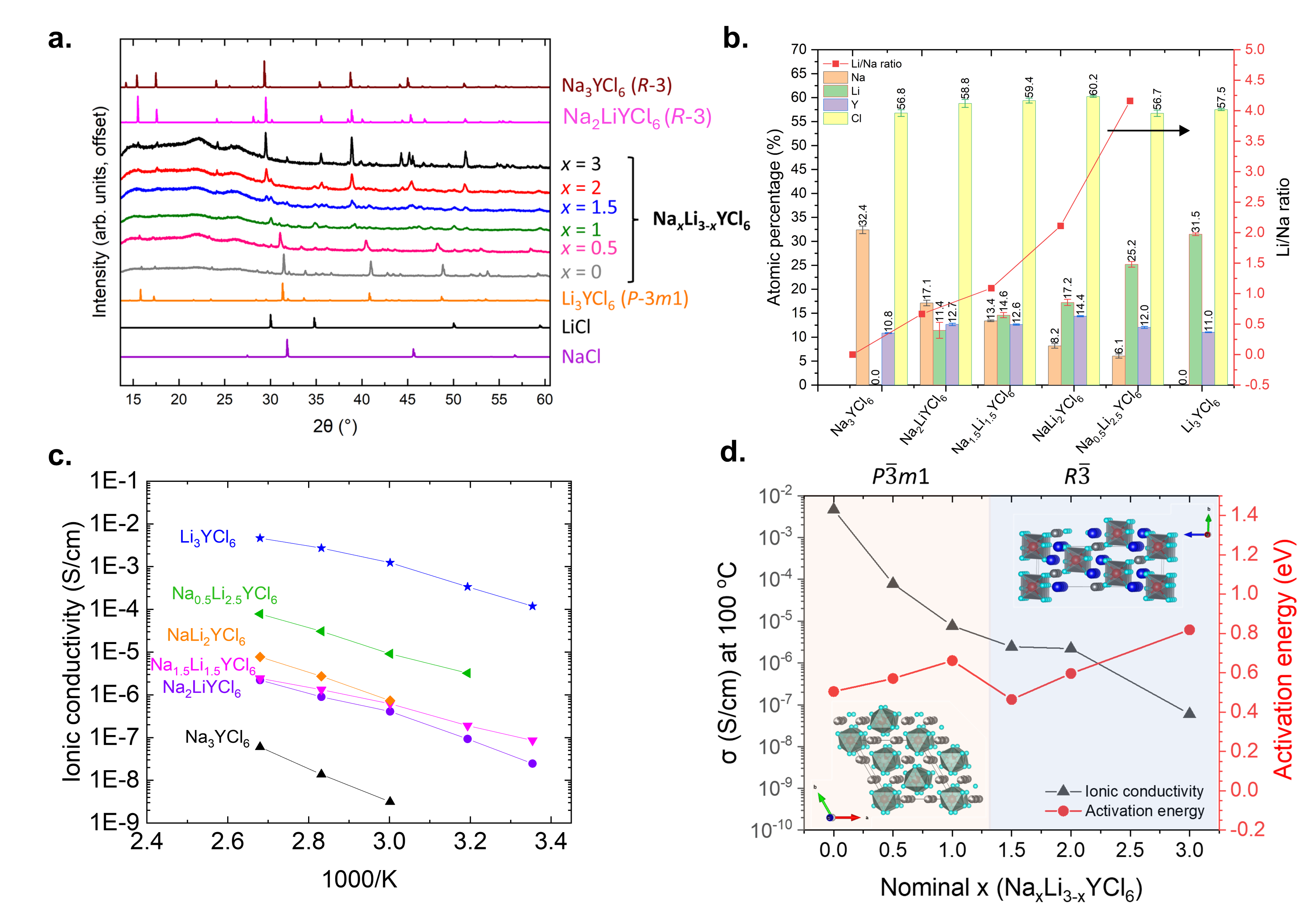}

    \caption{Experimental characterization data for the \naseries\ series.
    a.\ \Ac{PXRD} patterns of experimental samples with nominal compositions of \naseries\ ($x =$ 0, 0.5, 1, 1.5, 2, and 3) compared to calculated and reported patterns, including trigonal phases of Na$_3$YCl$_6$ and Li$_3$YCl$_6$.
    b.\ Atomic percentages of Na, Li, Y, Cl from \ac{XPS} analysis.
    c.\ Ionic conductivities of \naseries\ measured at different temperatures.
    d.\ Relationship between crystal structure and ionic conductivity.} \label{fig:series-characterization}
\end{figure}

The attempted synthetic approaches for Li$_3$YCl$_6$, Li$_5$YCl$_8$, and Li$_7$Y$_2$Cl$_{13}$ resulted in Li$_3$YCl$_6$ in the well-known P$\bar{3}$m1 phase~\cite{schlem_insights_2021} (see Supplementary Figure \ref{fig:pxrd-li3ycl6} to Supplementary Figure \ref{fig:pxrd-li7y2cl13}).
The \ac{PXRD} pattern of synthesized Na$_2$LiYCl$_6$ matched with the predicted structure in the Na$_3$YCl$_6$ R$\bar{3}$ phase, as shown in Figure~\ref{fig:series-characterization}.
Given the similarity between Li and Na, we also synthesized compositions of a \naseries\ series with different Na and Li ratios.
\Ac{PXRD} patterns (Figure~\ref{fig:series-characterization}a) reveal that the trigonal R$\bar{3}$ Na$_3$YCl$_6$ is the main phase observed for $x =$ 1.5 to 3~\cite{stenzel_ternare_1993},
while the main phase for $x =$ 0 to 0.5 samples matches the \ac{XRD} pattern of trigonal Li$_3$YCl$_6$ with space group P$\bar{3}$m1~\cite{schlem_insights_2021}.

Additional experiments are ongoing to fully understand the role of Na in these materials for chemical and structural stability.
Small amounts of unreacted NaCl and LiCl were observed in the synthesized $x = 3$ composition and $x =$ 1 to 2 compositions, respectively.
Sample crystallinity was greatly affected by humidity during the transfer and measurement of the samples to the diffractometer.
It was difficult to accurately refine unit cell parameters from \ac{PXRD} data due to large background from the Kapton tape and the broadened peaks of the \ac{PXRD} patterns.
Preliminary single-crystal \ac{XRD} data suggests successful synthesis of the Na$_2$LiYCl$_6$ R$\bar{3}$ phase.
Additional single-crystal \ac{XRD} measurements are ongoing to confirm successful synthesis for all compositions in this series.

\Ac{XPS} was used to confirm the atomic percentages of Li, Na, Y, and Cl in the \naseries\ samples.
For $x$ from 0 to 3, the Li/Na ratio matches qualitatively with the nominal values from the compositions (Figure~\ref{fig:series-characterization}b).
\Ac{SEM} and \ac{EDS} (see Figure~\ref{fig:eds-ratios} and Table~\ref{tab:elemental-data}) confirmed the relative elemental ratios of Na, Y, and Cl for the bulk of the samples.

\subsubsection{Conductivity}


The introduction of Li in place of Na remarkably boosted the ionic conductivity by more than two orders of magnitude, specifically at a substitution level of one third, exemplified by Na$_2$LiYCl$_6$ ($x=2$) exhibiting a conductivity of $2.2 \times 10^{−6}$ S/cm at 100~°C, compared to the meager $6.0 \times 10^{−8}$ S/cm at 100~°C observed in the parent Na$_3$YCl$_6$ material.
The measured ionic conductivities (Figure~\ref{fig:series-characterization}c) across varying $x$ values consistently depict the rising trend associated with increased Li content.
This trend strongly implies that the heightened mobility of Li$^+$ ions actively contributes to the measured conductivity within these substituted compositions.
Simultaneously, a significant reduction of activation energy (0.46--0.66 eV) was observed with the introduction of lithium ($x=$ 0.5 to 3) compared to the parent Na$_3$YCl$_6$ (0.82 eV).
Notably, Na$_2$LiYCl$_6$ (with a similar R$\bar{3}$ crystal structure as parent Na$_3$YCl$_6$, as shown in Figure \ref{fig:series-characterization}d) displays a significantly lower activation energy ($E_a =$ 0.6 eV), indicating an amplified ionic diffusion process.

This enhancement in conductivity and reduction in activation energy can likely be attributed to the involvement of Li$^+$ ions in ionic transportation, as well as potential alterations in the crystal structure, such as expansion of ion diffusion channels and local structural disorder resulting from Li$^+$ substitution.
The trigonal crystal structure of parent Na$_3$YCl$_6$ (R$\bar{3}$) contains one fully occupied and two partially occupied Na sites, enabling diffusion pathways through and between channels, predominantly facilitating ionic conduction along the $c$-axis \cite{qie_yttriumsodium_2020}.
The heightened conductivity observed upon lithium substitution suggests that the evolving occupancy and distribution of Na and Li sites potentially modulate the energy landscape along the $c$-axis, thereby fostering accelerated ionic diffusion.
Further investigation is warranted to validate the effects of Na/Li distribution on ion conduction pathways and the associated transport mechanisms.
Nonetheless, this material, Na$_x$Li$_{3−x}$YCl$_6$, discovered through our integrated computational approach, embodies a structural framework capable of accommodating two distinct mobile cations.
This characteristic not only signifies its potential for use in both Li- and Na-based solid-state batteries but also underscores its promise for designing new phases of ion-conducting materials through computational discovery.

\section{Discussion}

\textbf{Functional discovery.}
Discovering materials presents an enduring challenge, while identifying functional materials pertinent to specific applications proves even more formidable. 
Existing \ac{AI} models provide a good starting point in this process, but identifying and demonstrating the industry applicability of discovered materials requires accurate predictions and experimental validation, which can be time consuming, resource intensive, and expensive.
To expedite the discovery of industry-relevant materials, the initially large pool of material candidates must first be filtered, using \ac{AI} models, to produce a smaller pool of prospective candidates that are likely to be synthesizable. 
In this study, we have successfully used this approach, combining state-of-the-art \ac{AI} models with cloud \ac{HPC} to predict half a million stable and potentially synthesizable materials from an initial pool of over 32 million.
From that smaller pool of stable materials, the filters we applied revealed only a small number of functional materials with properties that would make them suitable for targeted applications.

\textbf{Designing solid-state electrolytes.}
We predicted new compositions of Na- and Li-containing electrolytes based on the known Na$_3$YCl$_6$ R$\bar{3}$ structure and we experimentally validated the formation of \naseries-based compounds.
The Na/Li substitutional compounds in this series showed experimental conductivities that make them viable candidates for future development as a solid-state electrolyte material.
The alkali cation substitution, resulting in stable halide phases exhibiting significant conductivity, represents a contemporary approach compared to the conventional aliovalent doping methodology reported for halide-based electrolytes \cite{wu_stable_2021, MEI2023232257}.
In particular, this structural framework's capability to host both Na and Li ions efficiently could pave the way for versatile battery designs, offering potential advantages in terms of cost, stability, and performance.
In addition to identifying these viable candidates, this discovery workflow also generated new computational data for guiding the design of electrolytes.
In particular, we rediscovered a decade's worth of \ac{ESW} trends for the widest range of materials to our knowledge.
The broader view of \ac{ESW} trends clearly suggests that the halide families have the most promising balance of \ac{ESW} and conductivity.
Our mechanical property predictions also suggest that these soft materials can form flexible interfaces with other battery components like solid electrodes, a main challenge in designing solid-state batteries.
We are currently assessing the electrochemical stability and performance of these newly discovered materials in solid-state battery systems.

\textbf{Next steps.}
The choice and execution of filters is very important when using computer-based methods to discover functional materials.
To focus our experimental work on only the most promising candidates, we applied a stringent \ac{ESW} filter that eliminated many material candidates in our search for a new solid electrolyte material.
Most known electrolytes, including sulfides such as Li$_{10}$GePS$_{12}$ and oxides such as Li$_7$La$_3$Zr$_2$O$_{12}$, would not have passed through the filters used in this study~\cite{han_electrochemical_2016}.
By relaxing the conditions of these filters, it is likely that we could discover additional electrolyte candidate materials among the 8,795 compounds that entered the \ac{ESW} filter. 
To train our \ac{ML} models, we used only materials with integer atomic occupancies that are conducive to quantum mechanical calculations.
However, it is very common for fast Li conductors to have disordered site occupancies and defects, neither of which are typically considered in high-throughput studies.
Additional work is needed to expand the materials space to consider more complex materials going beyond ordered, perfect crystals. 
Future optimization of the workflow beyond screening will be based on the direct generation of materials with given properties, such as the method recently demonstrated by MatterGen~\cite{zeni_mattergen_2023}.

\textbf{Broader impact.}
We demonstrated the integration of \ac{AI} and cloud \ac{HPC} as an end-to-end workflow, beginning with \ac{AI} exploration and enhanced by \ac{AI}/\ac{DFT} filters on cloud-based \ac{HPC}, which holds the potential to revolutionize functional materials discovery.
This transformative approach expands accessibility and significantly accelerates the identification, synthesis, and experimental validation of technologically relevant materials.
This fusion of techniques holds the potential to significantly reduce the time and resources spent on traditional trial-and-error experimental discovery of functional materials.

\section{Methods}

\subsection{\Ac{AI} model, training, and data}

We used the M3GNet model framework~\cite{chen_universal_2022} for training \ac{ML} potentials and retrained the Materials Project data on Standard\_ND40rs\_v2 \acp{VM}, each of which has 8 V100 32GB \acp{GPU}, on the Azure Quantum Elements platform. ML geometric optimization and inference were performed on Standard\_HC44rs \acp{VM}.

We used the Horovod~\cite{sergeev_horovod_2018} training framework on multi-\ac{GPU} training.
The training hyperparameters are consistent with previous work~\cite{chen_universal_2022}.
The other property models such as band gap and elasticity models are from previous work~\cite{chen_universal_2022}.

Materials Project data for training \ac{ML} potential were curated from the 2021.2.8 version and we expanded the structures from the previous M3GNet work.
Materials were selected only if they had a task\_type or task\_label field from Materials Project.
We filtered out materials that have energy per atom of less than -25 eV, or higher than 50 eV/atom, and we also removed structures that have an interatomic distance of less than 0.5 Å.

We curated a total of 117,970 materials, and for each material, we kept only the first, middle, and last relaxation snapshots in cases where the structure relaxation produced more than three snapshots.
This process produced a total of 352,767 structures, with 352,767 energy values, 30,148,593 force components, and 2,116,602 unique stress components.
We then divided the materials into training, validation, and testing categories.
The snapshots in each category were 317,489, 17,633, and 17,645, respectively.
The model error for the test data was 29.9 meV/atom, 72.1 meV/Å, and 0.40 GPa.

\subsection{\Ac{DFT}}

\Ac{DFT} and \ac{AIMD} calculations were performed with VASP~\cite{kresse_ab_1993,kresse_efficiency_1996}.
For structural relaxation, static calculations, and band-structure calculations, we used the Atomate workflows~\cite{mathew_atomate_2017}, FireWorks~\cite{jain_fireworks_2015} workflow manager, and custodian error handlers~\cite{ong_python_2013}, which is consistent with settings used in the Materials Project~\cite{jain_commentary_2013}.
All DFT calculations were run on the Azure Quantum Elements platform using the Standard\_HB120rs\_v2 120-core VMs.

For \ac{AIMD}, we created super-cells of at least 80 atoms. 
We used a time step of 2 fs and ran 100,000 steps (total 200 ps) for each \ac{MD} trajectory at temperatures from 400 K to 700 K with `MPMDSet' in pymatgen~\cite{ong_python_2013}.  For Na$_2$LiYCl$_6$, which has a lower conductivity than other candidates, we raised the simulation temperature range to 700 K to 1000 K. The materials did not melt, which demonstrates their stability. We used Niggli reduction using Pymatgen\cite{ong_python_2013} on the structure to obtain more regular lattices for simulations.

\subsection{Synthesis methods}

All reagents of sodium chloride (Sigma Aldrich, 99.999\%), lithium chloride (Sigma Aldrich, anhydrous, 99.95\%), and yttrium chloride (Sigma Aldrich, anhydrous, 99.99\%) were used as received.
All material handling of reagents and synthesized samples was carried out in an Ar-filled glovebox ($<0.5$ ppm O$_2$, 0.5 ppm moisture) to minimize water contamination.
Stoichiometric ratios of the NaCl, LiCl, and YCl$_3$ powders were weighed out, ground together via mortar-and-pestle hand grinding, and then compacted into a pellet via a hydraulic press (200 bar) in stainless steel pressure dies.
Each pellet was placed into a quartz ampoule, evacuated, flame-sealed, and then placed into a tube furnace with the pellet sample nearest to the thermocouple.
The annealing temperature varied between 450~°C and 550~°C depending on the stoichiometry of the sample.
The temperature profile used was a 10~°C/min ramp to the respective temperature followed by annealing for 5 hours.
The furnace was then turned off to allow the samples to cool naturally.
Once cooled, the sample ampoules were carefully opened in a glovebox, finely ground with a mortar and pestle, and then characterized.
All products remained in the glovebox while handling.

\subsection{Characterization methods}

\subsubsection{\Ac{XRD}}

All powder X-ray diffraction measurements were conducted on a Rigaku SmartLab SE Bragg-Brentano diffractometer with a Cu source ($\lambda = 1.5418$ Å) and D/teX Ultra 250 1D high-speed position-sensitive detector.
Samples were placed in a SiO$_2$ zero background well holder then covered with a single layer of Kapton tape inside a glovebox to avoid moisture degradation.

\subsubsection{\Ac{XPS}}

The \naseries\ (100 mg) fine powder was compacted into a pellet via a hydraulic press (100 bar) in stainless steel pressure dies inside an Ar-filled glovebox.
The pellet was then loaded onto a sample holder and transferred to an \ac{XPS} analysis chamber without exposing to air.
\Ac{XPS} analysis was performed using a Kratos Axis Ultra DLD spectrometer, which consists of an Al K$\alpha$ monochromatic X-ray source (1486.6 eV) and a high-resolution spherical mirror analyzer.
The X-ray source was operated at 150 W power and the emitted photoelectrons were collected at the analyzer entrance slit normal to the sample surface.
The high-resolution spectra were collected at a pass energy of 40 eV with a step size of 0.1 eV.
\Ac{XPS} spectra were calibrated using C 1s signal at 285 eV.
Data were  processed using CasaXPS software.
The Li 1s, Na 2s, Y 3d, and Cl 2p high-resolution spectra were used to calculate the composition of the samples.

\subsubsection{\Ac{SEM} and \ac{EDS}}

To confirm the relative ratios of Na, Y, and Cl in the samples, elemental analysis was performed using \ac{SEM} and \ac{EDS}.
Li is a light element so it cannot be detected by \ac{EDS}.
Therefore, only relative ratios of the non-Li elements were determined via this technique.
Powders of sample were pressed into a flat, thin pellet and placed onto a conductive double-sided carbon adhesive dot placed on an aluminum stub.
An electron beam at 20 kV accelerating voltage with 1.6 nA current was used with a working distance of 10 mm.
At least 10 different sites on each pellet of different compositions (\naseries; $x$ = 0, 0.5, 1, 1.5, 2, 3) were measured to generate average statistics of the elemental quantification.
Oxygen content was noted on the surface despite the quick transfer from the glovebox to the \ac{SEM} instrument.

\subsubsection{\Ac{EIS} measurement}

To determine ionic conductivity, 110 mg of fine electrolyte powder was loaded into a Split Cell (MSE Supplies) with a diameter of 12 mm, sandwiched between 2 stainless steel electrodes.
The cell was then cold-pressed at 100 bars using a hydraulic press inside an Ar-filled glovebox.
The cell was then removed from the glovebox, loaded into a pressure jig, and pressurized to 250 MPa. Impedance measurements were taken with an applied AC potential of 100 mV over a frequency range of 1 MHz to 100 mHz using Biologic VMP3.
The \ac{EIS} were collected at various temperatures between 100~°C and 25~°C. At each temperature, the cell was rested for 2 h prior to collecting the \ac{EIS}.
The thickness of each pellet was measured using a digital micrometer (Mitutoyo) after completing the impedance measurement.
The ionic conductivity of the pelletized electrolyte was determined using the \ac{EIS} data.

\section*{Data and code availability}
Code is available at \url{https://github.com/materialsvirtuallab/m3gnet}.

\section*{Acknowledgments}

The authors thank the rest of the Microsoft Azure Quantum team for their support on engineering infrastructure.
The authors thank Deborah Hutchinson for paper revisions.
A portion of this research was performed on a project award DOI:\href{https://doi.org/10.46936/prop.proj.2023.61032/60012343}{10.46936/prop.proj.2023.61032/60012343} from the Environmental Molecular Sciences Laboratory, a DOE Office of Science User Facility sponsored by the Biological and Environmental Research program under Contract No.\ DE-AC05-76RL01830.
The experimental research work conducted at PNNL was funded by Microsoft.

\section*{Author contributions}

C.C.\ and M.T.\ conceived of the project.
C.C.\ designed the workflow and carried out the calculations with input from N.B.\ and M.T.
The project was managed by L.L., N.B., and B.B.\ who also procured resources.
V.M. and D.N.\ conceived and designed the experimental analysis.
D.N., A.K., and S.L.\ collected the experimental data.
V.M., D.N., A.K., S.L., and K.M.\ analyzed and interpreted the experimental data.
C.C., N.B., D.N., and S.L.\ wrote the paper.
All authors contributed to discussions, analysis, and paper revision.

\section*{Competing interests}

C.C., N.B, L.L., C.O., B.B., and M.T.\ work at Microsoft.
Two provisional patents were filed.

\clearpage


\bibliography{sn-bibliography}

\begin{thebibliography}{10}
\expandafter\ifx\csname url\endcsname\relax
  \def\url#1{\burl{#1}}\fi
\expandafter\ifx\csname urlprefix\endcsname\relax\def\urlprefix{URL }\fi
\providecommand{\bibinfo}[2]{#2}
\providecommand{\eprint}[2][]{\url{#2}}
\providecommand{\doi}[1]{\url{https://doi.org/#1}}
\bibcommenthead

\bibitem{jain_commentary_2013}
\bibinfo{author}{Jain, A.} \emph{et~al.}
\newblock \bibinfo{title}{Commentary: {The} {Materials} {Project}: {A} materials genome approach to accelerating materials innovation}.
\newblock \emph{\bibinfo{journal}{APL Materials}} \textbf{\bibinfo{volume}{1}}, \bibinfo{pages}{011002} (\bibinfo{year}{2013}).
\newblock \urlprefix\url{https://doi.org/10.1063/1.4812323}.

\bibitem{curtarolo_aflowliborg_2012}
\bibinfo{author}{Curtarolo, S.} \emph{et~al.}
\newblock \bibinfo{title}{{AFLOWLIB}.{ORG}: {A} distributed materials properties repository from high-throughput ab initio calculations}.
\newblock \emph{\bibinfo{journal}{Computational Materials Science}} \textbf{\bibinfo{volume}{58}}, \bibinfo{pages}{227--235} (\bibinfo{year}{2012}).
\newblock \urlprefix\url{https://doi.org/10.1016/j.commatsci.2012.02.002}.

\bibitem{saal_materials_2013}
\bibinfo{author}{Saal, J.~E.}, \bibinfo{author}{Kirklin, S.}, \bibinfo{author}{Aykol, M.}, \bibinfo{author}{Meredig, B.} \& \bibinfo{author}{Wolverton, C.}
\newblock \bibinfo{title}{Materials design and discovery with high-throughput density functional theory: {The} {Open} {Quantum} {Materials} {Database} ({OQMD})}.
\newblock \emph{\bibinfo{journal}{JOM}} \textbf{\bibinfo{volume}{65}}, \bibinfo{pages}{1501--1509} (\bibinfo{year}{2013}).
\newblock \urlprefix\url{https://doi.org/10.1007/s11837-013-0755-4}.

\bibitem{talirz_materials_2020}
\bibinfo{author}{Talirz, L.} \emph{et~al.}
\newblock \bibinfo{title}{Materials {Cloud}, a platform for open computational science}.
\newblock \emph{\bibinfo{journal}{Scientific Data}} \textbf{\bibinfo{volume}{7}}, \bibinfo{pages}{299} (\bibinfo{year}{2020}).
\newblock \urlprefix\url{https://doi.org/10.1038/s41597-020-00637-5}.

\bibitem{chen_graph_2019}
\bibinfo{author}{Chen, C.}, \bibinfo{author}{Ye, W.}, \bibinfo{author}{Zuo, Y.}, \bibinfo{author}{Zheng, C.} \& \bibinfo{author}{Ong, S.~P.}
\newblock \bibinfo{title}{Graph networks as a universal machine learning framework for molecules and crystals}.
\newblock \emph{\bibinfo{journal}{Chemistry of Materials}} \textbf{\bibinfo{volume}{31}}, \bibinfo{pages}{3564--3572} (\bibinfo{year}{2019}).
\newblock \urlprefix\url{https://doi.org/10.1021/acs.chemmater.9b01294}.

\bibitem{xie_crystal_2018}
\bibinfo{author}{Xie, T.} \& \bibinfo{author}{Grossman, J.~C.}
\newblock \bibinfo{title}{Crystal graph convolutional neural networks for an accurate and interpretable prediction of material properties}.
\newblock \emph{\bibinfo{journal}{Physical Review Letters}} \textbf{\bibinfo{volume}{120}}, \bibinfo{pages}{145301} (\bibinfo{year}{2018}).
\newblock \urlprefix\url{https://doi.org/10.1103/PhysRevLett.120.145301}.

\bibitem{choudhary_atomistic_2021}
\bibinfo{author}{Choudhary, K.} \& \bibinfo{author}{DeCost, B.}
\newblock \bibinfo{title}{Atomistic line graph neural network for improved materials property predictions}.
\newblock \emph{\bibinfo{journal}{npj Computational Materials}} \textbf{\bibinfo{volume}{7}}, \bibinfo{pages}{1--8} (\bibinfo{year}{2021}).
\newblock \urlprefix\url{https://doi.org/10.1038/s41524-021-00650-1}.

\bibitem{chen_universal_2022}
\bibinfo{author}{Chen, C.} \& \bibinfo{author}{Ong, S.~P.}
\newblock \bibinfo{title}{A universal graph deep learning interatomic potential for the periodic table}.
\newblock \emph{\bibinfo{journal}{Nature Computational Science}} \textbf{\bibinfo{volume}{2}}, \bibinfo{pages}{718--728} (\bibinfo{year}{2022}).
\newblock \urlprefix\url{https://doi.org/10.1038/s43588-022-00349-3}.

\bibitem{deng_chgnet_2023}
\bibinfo{author}{Deng, B.} \emph{et~al.}
\newblock \bibinfo{title}{{CHGNet} as a pretrained universal neural network potential for charge-informed atomistic modelling}.
\newblock \emph{\bibinfo{journal}{Nature Machine Intelligence}} \textbf{\bibinfo{volume}{5}}, \bibinfo{pages}{1031--1041} (\bibinfo{year}{2023}).
\newblock \urlprefix\url{https://www.nature.com/articles/s42256-023-00716-3}.
\newblock \bibinfo{note}{Number: 9 Publisher: Nature Publishing Group}.

\bibitem{merchant_scaling_2023}
\bibinfo{author}{Merchant, A.} \emph{et~al.}
\newblock \bibinfo{title}{Scaling deep learning for materials discovery}.
\newblock \emph{\bibinfo{journal}{Nature}} \bibinfo{pages}{1--6} (\bibinfo{year}{2023}).
\newblock \urlprefix\url{https://doi.org/10.1038/s41586-023-06735-9}.

\bibitem{szymanski_autonomous_2023}
\bibinfo{author}{Szymanski, N.~J.} \emph{et~al.}
\newblock \bibinfo{title}{An autonomous laboratory for the accelerated synthesis of novel materials}.
\newblock \emph{\bibinfo{journal}{Nature}} \bibinfo{pages}{1--6} (\bibinfo{year}{2023}).
\newblock \urlprefix\url{https://doi.org/10.1038/s41586-023-06734-w}.

\bibitem{zunger_inverse_2018}
\bibinfo{author}{Zunger, A.}
\newblock \bibinfo{title}{Inverse design in search of materials with target functionalities}.
\newblock \emph{\bibinfo{journal}{Nature Reviews Chemistry}} \textbf{\bibinfo{volume}{2}}, \bibinfo{pages}{1--16} (\bibinfo{year}{2018}).
\newblock \urlprefix\url{https://doi.org/10.1038/s41570-018-0121}.

\bibitem{riebesell_pymatviz_2022}
\bibinfo{author}{Riebesell, J.}
\newblock \bibinfo{title}{{Pymatviz}: visualization toolkit for materials informatics} (\bibinfo{year}{2022}).
\newblock \urlprefix\url{https://github.com/janosh/pymatviz}.

\bibitem{hautier_data_2011}
\bibinfo{author}{Hautier, G.}, \bibinfo{author}{Fischer, C.}, \bibinfo{author}{Ehrlacher, V.}, \bibinfo{author}{Jain, A.} \& \bibinfo{author}{Ceder, G.}
\newblock \bibinfo{title}{Data mined ionic substitutions for the discovery of new compounds}.
\newblock \emph{\bibinfo{journal}{Inorganic Chemistry}} \textbf{\bibinfo{volume}{50}}, \bibinfo{pages}{656--663} (\bibinfo{year}{2011}).
\newblock \urlprefix\url{https://doi.org/10.1021/ic102031h}.

\bibitem{wu_first_2013}
\bibinfo{author}{Wu, Y.}, \bibinfo{author}{Lazic, P.}, \bibinfo{author}{Hautier, G.}, \bibinfo{author}{Persson, K.} \& \bibinfo{author}{Ceder, G.}
\newblock \bibinfo{title}{First principles high throughput screening of oxynitrides for water-splitting photocatalysts}.
\newblock \emph{\bibinfo{journal}{Energy \& Environmental Science}} \textbf{\bibinfo{volume}{6}}, \bibinfo{pages}{157--168} (\bibinfo{year}{2013}).
\newblock \urlprefix\url{https://pubs.rsc.org/en/content/articlelanding/2013/ee/c2ee23482c}.

\bibitem{wang_mining_2018}
\bibinfo{author}{Wang, Z.} \emph{et~al.}
\newblock \bibinfo{title}{Mining unexplored chemistries for phosphors for high-color-quality white-light-emitting diodes}.
\newblock \emph{\bibinfo{journal}{Joule}} \textbf{\bibinfo{volume}{2}}, \bibinfo{pages}{914--926} (\bibinfo{year}{2018}).
\newblock \urlprefix\url{https://www.sciencedirect.com/science/article/pii/S2542435118300436}.

\bibitem{zhu_li3yps42_2017}
\bibinfo{author}{Zhu, Z.}, \bibinfo{author}{Chu, I.-H.} \& \bibinfo{author}{Ong, S.~P.}
\newblock \bibinfo{title}{{Li$_3$Y(PS$_4$)$_2$} and {Li$_5$PS$_4$Cl$_2$}: New lithium superionic conductors predicted from silver thiophosphates using efficiently tiered ab initio molecular dynamics simulations}.
\newblock \emph{\bibinfo{journal}{Chemistry of Materials}} \textbf{\bibinfo{volume}{29}}, \bibinfo{pages}{2474--2484} (\bibinfo{year}{2017}).
\newblock \urlprefix\url{https://doi.org/10.1021/acs.chemmater.6b04049}.

\bibitem{ong_python_2013}
\bibinfo{author}{Ong, S.~P.} \emph{et~al.}
\newblock \bibinfo{title}{Python {Materials} {Genomics} (pymatgen): A robust, open-source python library for materials analysis}.
\newblock \emph{\bibinfo{journal}{Computational Materials Science}} \textbf{\bibinfo{volume}{68}}, \bibinfo{pages}{314--319} (\bibinfo{year}{2013}).
\newblock \urlprefix\url{https://doi.org/10.1016/j.commatsci.2012.10.028}.

\bibitem{belsky_new_2002}
\bibinfo{author}{Belsky, A.}, \bibinfo{author}{Hellenbrandt, M.}, \bibinfo{author}{Karen, V.~L.} \& \bibinfo{author}{Luksch, P.}
\newblock \bibinfo{title}{New developments in the {Inorganic} {Crystal} {Structure} {Database} ({ICSD}): accessibility in support of materials research and design}.
\newblock \emph{\bibinfo{journal}{Acta Crystallographica Section B: Structural Science}} \textbf{\bibinfo{volume}{58}}, \bibinfo{pages}{364--369} (\bibinfo{year}{2002}).
\newblock \urlprefix\url{https://doi.org/10.1107/S0108768102006948}.

\bibitem{pickard_high-pressure_2006}
\bibinfo{author}{Pickard, C.~J.} \& \bibinfo{author}{Needs, R.~J.}
\newblock \bibinfo{title}{High-pressure phases of silane}.
\newblock \emph{\bibinfo{journal}{Physical Review Letters}} \textbf{\bibinfo{volume}{97}}, \bibinfo{pages}{045504} (\bibinfo{year}{2006}).
\newblock \urlprefix\url{https://doi.org/10.1103/PhysRevLett.97.045504}.

\bibitem{glass_uspexevolutionary_2006}
\bibinfo{author}{Glass, C.~W.}, \bibinfo{author}{Oganov, A.~R.} \& \bibinfo{author}{Hansen, N.}
\newblock \bibinfo{title}{{USPEX}—{Evolutionary} crystal structure prediction}.
\newblock \emph{\bibinfo{journal}{Computer Physics Communications}} \textbf{\bibinfo{volume}{175}}, \bibinfo{pages}{713--720} (\bibinfo{year}{2006}).
\newblock \urlprefix\url{https://doi.org/10.1016/j.cpc.2006.07.020}.

\bibitem{wang_calypso_2012}
\bibinfo{author}{Wang, Y.}, \bibinfo{author}{Lv, J.}, \bibinfo{author}{Zhu, L.} \& \bibinfo{author}{Ma, Y.}
\newblock \bibinfo{title}{{CALYPSO}: {A} method for crystal structure prediction}.
\newblock \emph{\bibinfo{journal}{Computer Physics Communications}} \textbf{\bibinfo{volume}{183}}, \bibinfo{pages}{2063--2070} (\bibinfo{year}{2012}).
\newblock \urlprefix\url{https://doi.org/10.1016/j.cpc.2012.05.008}.

\bibitem{azhop}
\bibinfo{title}{Azure {HPC} {On-Demand} {Platform}}.
\newblock \urlprefix\url{https://azure.github.io/az-hop/}.
\newblock \bibinfo{note}{Accessed 2023}.

\bibitem{aqe}
\bibinfo{title}{Azure {Quantum} {Elements}}.
\newblock \urlprefix\url{https://quantum.microsoft.com/en-us/our-story/quantum-elements-overview}.
\newblock \bibinfo{note}{Accessed 2023}.

\bibitem{docker}
\bibinfo{title}{Docker}.
\newblock \urlprefix\url{https://www.docker.com}.
\newblock \bibinfo{note}{Accessed 2023}.

\bibitem{netapp}
\bibinfo{title}{{NetApp} {Files}}.
\newblock \urlprefix\url{https://www.netapp.com/}.
\newblock \bibinfo{note}{Accessed 2023}.

\bibitem{slurm}
\bibinfo{title}{{Simple} {Linux} {Utility} for {Resource} {Management} {(SLURM)}}.
\newblock \urlprefix\url{https://slurm.schedmd.com/overview.html}.
\newblock \bibinfo{note}{Accessed 2023}.

\bibitem{cosmosdb}
\bibinfo{title}{{CosmosDB}}.
\newblock \urlprefix\url{https://azure.microsoft.com/en-us/products/cosmos-db/}.
\newblock \bibinfo{note}{Accessed 2023}.

\bibitem{mongodb_api}
\bibinfo{title}{{MongoDB} query {API}}.
\newblock \urlprefix\url{https://www.mongodb.com/docs/manual/query-api/}.
\newblock \bibinfo{note}{Accessed 2023}.

\bibitem{wang_lithium_2019}
\bibinfo{author}{Wang, S.} \emph{et~al.}
\newblock \bibinfo{title}{Lithium chlorides and bromides as promising solid-state chemistries for fast ion conductors with good electrochemical stability}.
\newblock \emph{\bibinfo{journal}{Angewandte Chemie International Edition}} \textbf{\bibinfo{volume}{58}}, \bibinfo{pages}{8039--8043} (\bibinfo{year}{2019}).
\newblock \urlprefix\url{https://doi.org/10.1002/anie.201901938}.

\bibitem{tan_synthesis_2011}
\bibinfo{author}{Tan, J.} \& \bibinfo{author}{Tiwari, A.}
\newblock \bibinfo{title}{Synthesis of cubic phase {Li$_7$La$_3$Zr$_2$O$_{12}$} electrolyte for solid-state lithium-ion batteries}.
\newblock \emph{\bibinfo{journal}{Electrochemical and Solid-State Letters}} \textbf{\bibinfo{volume}{15}}, \bibinfo{pages}{A37} (\bibinfo{year}{2011}).
\newblock \urlprefix\url{https://doi.org/10.1149/2.003203esl}.

\bibitem{schlem_insights_2021}
\bibinfo{author}{Schlem, R.}, \bibinfo{author}{Banik, A.}, \bibinfo{author}{Ohno, S.}, \bibinfo{author}{Suard, E.} \& \bibinfo{author}{Zeier, W.~G.}
\newblock \bibinfo{title}{Insights into the lithium sub-structure of superionic conductors {Li$_3$YCl$_6$} and {Li$_3$YBr$_6$}}.
\newblock \emph{\bibinfo{journal}{Chemistry of Materials}} \textbf{\bibinfo{volume}{33}}, \bibinfo{pages}{327--337} (\bibinfo{year}{2021}).
\newblock \urlprefix\url{https://doi.org/10.1021/acs.chemmater.0c04352}.

\bibitem{asano_solid_2018}
\bibinfo{author}{Asano, T.} \emph{et~al.}
\newblock \bibinfo{title}{Solid halide electrolytes with high lithium-ion conductivity for application in 4 {V} class bulk-type all-solid-state batteries}.
\newblock \emph{\bibinfo{journal}{Advanced Materials}} \textbf{\bibinfo{volume}{30}}, \bibinfo{pages}{1803075} (\bibinfo{year}{2018}).
\newblock \urlprefix\url{https://doi.org/10.1002/adma.201803075}.

\bibitem{hu_revealing_2023}
\bibinfo{author}{Hu, L.} \emph{et~al.}
\newblock \bibinfo{title}{Revealing the {Pnma} crystal structure and ion-transport mechanism of the {Li$_3$YCl$_6$} solid electrolyte}.
\newblock \emph{\bibinfo{journal}{Cell Reports Physical Science}} \textbf{\bibinfo{volume}{4}}, \bibinfo{pages}{101428} (\bibinfo{year}{2023}).
\newblock \urlprefix\url{https://doi.org/10.1016/j.xcrp.2023.101428}.

\bibitem{steiner_neue_1992}
\bibinfo{author}{Steiner, H.-J.} \& \bibinfo{author}{Lutz, H.~D.}
\newblock \bibinfo{title}{Neue schnelle ionenleiter vom typ {MMIIICl6} ({MI} = {Li}, {Na}, {Ag}; {MIII} = {In}, {Y})}.
\newblock \emph{\bibinfo{journal}{Zeitschrift für anorganische und allgemeine Chemie}} \textbf{\bibinfo{volume}{613}}, \bibinfo{pages}{26--30} (\bibinfo{year}{1992}).
\newblock \urlprefix\url{https://doi.org/10.1002/zaac.19926130104}.

\bibitem{stenzel_ternare_1993}
\bibinfo{author}{Stenzel, F.} \& \bibinfo{author}{Meyer, G.}
\newblock \bibinfo{title}{Ternare halogenide vom typ {A$_3$MX$_6$}. {II}. {Das} system {Ag$_{3-x}$Na$_x$YCl$_6$}: Synthese, strukturen, ionenleitfahigkeit}.
\newblock \emph{\bibinfo{journal}{Zeitschrift für anorganische und allgemeine Chemie}} \textbf{\bibinfo{volume}{619}}, \bibinfo{pages}{652--660} (\bibinfo{year}{1993}).
\newblock \urlprefix\url{https://doi.org/10.1002/zaac.19936190406}.

\bibitem{qie_yttriumsodium_2020}
\bibinfo{author}{Qie, Y.} \emph{et~al.}
\newblock \bibinfo{title}{Yttrium–sodium halides as promising solid-state electrolytes with high ionic conductivity and stability for {Na}-ion batteries}.
\newblock \emph{\bibinfo{journal}{The Journal of Physical Chemistry Letters}} \textbf{\bibinfo{volume}{11}}, \bibinfo{pages}{3376--3383} (\bibinfo{year}{2020}).
\newblock \urlprefix\url{https://doi.org/10.1021/acs.jpclett.0c00010}.

\bibitem{wu_stable_2021}
\bibinfo{author}{Wu, E.~A.} \emph{et~al.}
\newblock \bibinfo{title}{A stable cathode-solid electrolyte composite for high-voltage, long-cycle-life solid-state sodium-ion batteries}.
\newblock \emph{\bibinfo{journal}{Nature Communications}} \textbf{\bibinfo{volume}{12}}, \bibinfo{pages}{1256} (\bibinfo{year}{2021}).
\newblock \urlprefix\url{https://doi.org/10.1038/s41467-021-21488-7}.

\bibitem{MEI2023232257}
\bibinfo{author}{Mei, H.-X.}, \bibinfo{author}{Piccardo, P.}, \bibinfo{author}{Cingolani, A.} \& \bibinfo{author}{Spotorno, R.}
\newblock \bibinfo{title}{Unconventional solid-state electrolytes for lithium-based batteries: Recent advances and challenges}.
\newblock \emph{\bibinfo{journal}{Journal of Power Sources}} \textbf{\bibinfo{volume}{553}}, \bibinfo{pages}{232257} (\bibinfo{year}{2023}).
\newblock \urlprefix\url{https://www.sciencedirect.com/science/article/pii/S0378775322012344}.

\bibitem{han_electrochemical_2016}
\bibinfo{author}{Han, F.}, \bibinfo{author}{Zhu, Y.}, \bibinfo{author}{He, X.}, \bibinfo{author}{Mo, Y.} \& \bibinfo{author}{Wang, C.}
\newblock \bibinfo{title}{Electrochemical stability of {Li$_{10}$GeP$_2$S$_{12}$} and {Li$_7$La$_3$Zr$_2$O$_{12}$} solid electrolytes}.
\newblock \emph{\bibinfo{journal}{Advanced Energy Materials}} \textbf{\bibinfo{volume}{6}} (\bibinfo{year}{2016}).
\newblock \urlprefix\url{https://doi.org/10.1002/aenm.201501590}.

\bibitem{zeni_mattergen_2023}
\bibinfo{author}{Zeni, C.} \emph{et~al.}
\newblock \bibinfo{title}{{MatterGen}: a generative model for inorganic materials design} (\bibinfo{year}{2023}).
\newblock \urlprefix\url{http://doi.org/10.48550/arXiv.2312.03687}.

\bibitem{sergeev_horovod_2018}
\bibinfo{author}{Sergeev, A.} \& \bibinfo{author}{Del~Balso, M.}
\newblock \bibinfo{title}{Horovod: fast and easy distributed deep learning in {TensorFlow}} (\bibinfo{year}{2018}).
\newblock \urlprefix\url{http://doi.org/10.48550/arXiv.1802.05799}.

\bibitem{kresse_ab_1993}
\bibinfo{author}{Kresse, G.} \& \bibinfo{author}{Hafner, J.}
\newblock \bibinfo{title}{Ab initio molecular dynamics for liquid metals}.
\newblock \emph{\bibinfo{journal}{Physical Review B}} \textbf{\bibinfo{volume}{47}}, \bibinfo{pages}{558--561} (\bibinfo{year}{1993}).
\newblock \urlprefix\url{https://doi.org/10.1103/PhysRevB.47.558}.

\bibitem{kresse_efficiency_1996}
\bibinfo{author}{Kresse, G.} \& \bibinfo{author}{Furthmüller, J.}
\newblock \bibinfo{title}{Efficiency of ab-initio total energy calculations for metals and semiconductors using a plane-wave basis set}.
\newblock \emph{\bibinfo{journal}{Computational Materials Science}} \textbf{\bibinfo{volume}{6}}, \bibinfo{pages}{15--50} (\bibinfo{year}{1996}).
\newblock \urlprefix\url{https://doi.org/10.1016/0927-0256(96)00008-0}.

\bibitem{mathew_atomate_2017}
\bibinfo{author}{Mathew, K.} \emph{et~al.}
\newblock \bibinfo{title}{Atomate: A high-level interface to generate, execute, and analyze computational materials science workflows}.
\newblock \emph{\bibinfo{journal}{Computational Materials Science}} \textbf{\bibinfo{volume}{139}}, \bibinfo{pages}{140--152} (\bibinfo{year}{2017}).
\newblock \urlprefix\url{https://doi.org/10.1016/j.commatsci.2017.07.030}.

\bibitem{jain_fireworks_2015}
\bibinfo{author}{Jain, A.} \emph{et~al.}
\newblock \bibinfo{title}{{FireWorks}: a dynamic workflow system designed for high-throughput applications}.
\newblock \emph{\bibinfo{journal}{Concurrency and Computation: Practice and Experience}} \textbf{\bibinfo{volume}{27}}, \bibinfo{pages}{5037--5059} (\bibinfo{year}{2015}).
\newblock \urlprefix\url{https://doi.org/10.1002/cpe.3505}.

\end{thebibliography}

\clearpage

\begin{appendices}

\renewcommand\thefigure{S\arabic{figure}}
\setcounter{figure}{0}
\renewcommand\thetable{S\arabic{table}}
\setcounter{table}{0}

\section{Acronyms}


\begin{acronym}[PXRD]
    \acro{AI}{artificial intelligence}
    \acro{AIMD}{\textit{ab initio} molecular dynamics}
    \acro{DFT}{density functional theory}
    \acro{EDS}{energy-dispersive X-ray spectroscopy}
    \acro{EIS}{electrochemical impedance spectra}
    \acro{ESW}{electrochemical stability window}
    \acro{HPC}{high-performance computing}
    \acro{GPU}{graphics processing unit}
    \acro{ICSD}{Inorganic Crystal Structure Database}
    \acro{MD}{molecular dynamics}
    \acro{ML}{machine learning}
    \acro{PBE}{Perdew-Burke-Ernzerhof}
    \acro{PXRD}{powder X-ray diffraction}
    \acro{SEM}{scanning electron microscopy}
    \acro{VM}{virtual machine}
    \acro{XPS}{X-ray photoelectron spectroscopy}
    \acro{XRD}{X-ray diffraction}
\end{acronym}

\section{Supplementary information}

\begin{figure}[hb]
    \centering
    \includegraphics[width=0.95\textwidth]{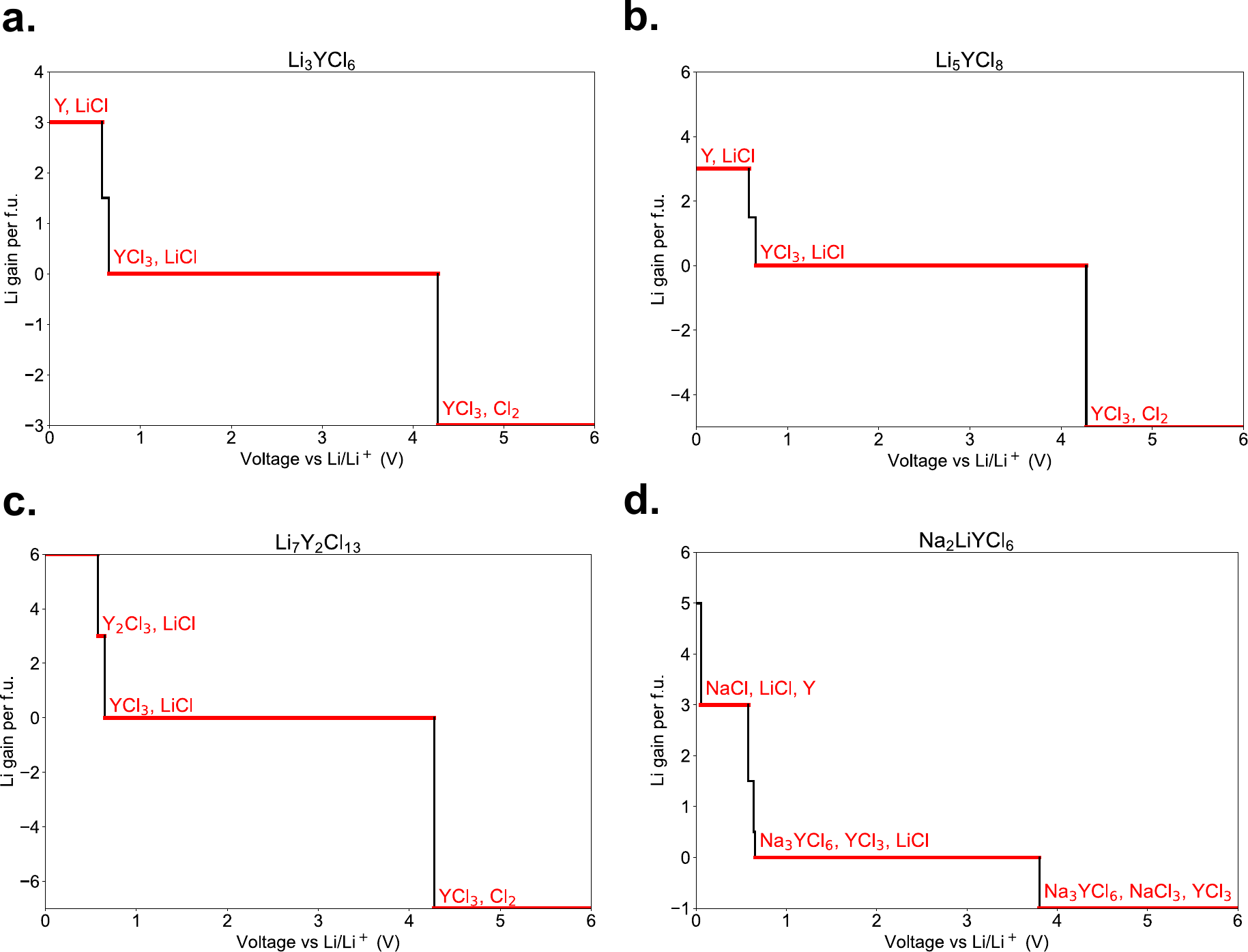}
    \caption{Li uptake and loss from reactions as a function of voltage.
    The voltage window where no Li uptake or loss occurred is defined as the electrochemical stability window.} \label{fig:top-esw}
\end{figure}

\begin{figure}
    \centering
    a.\ \includegraphics[width=0.45\textwidth]{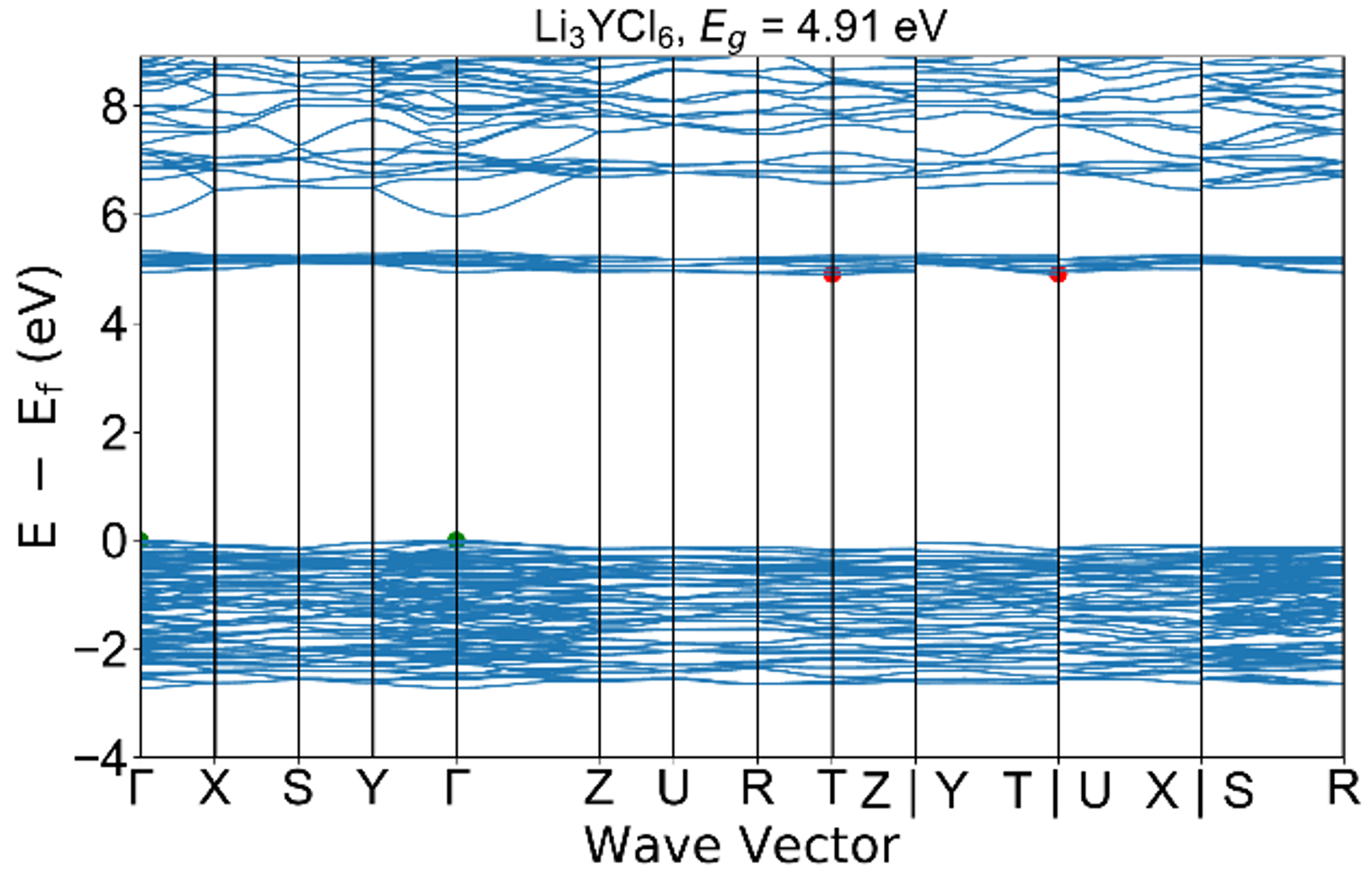}
    b.\ \includegraphics[width=0.45\textwidth]{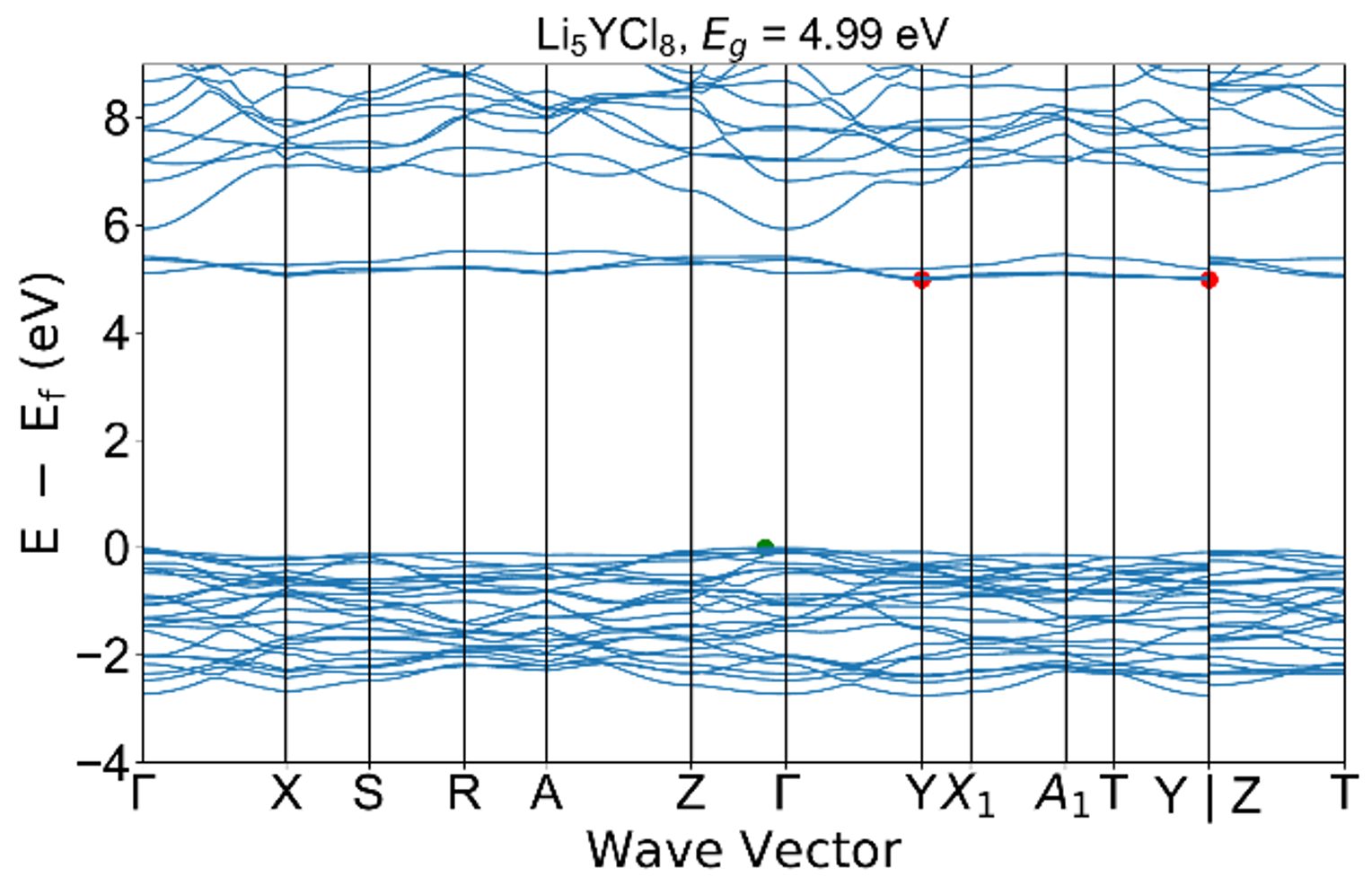}
    c.\ \includegraphics[width=0.45\textwidth]{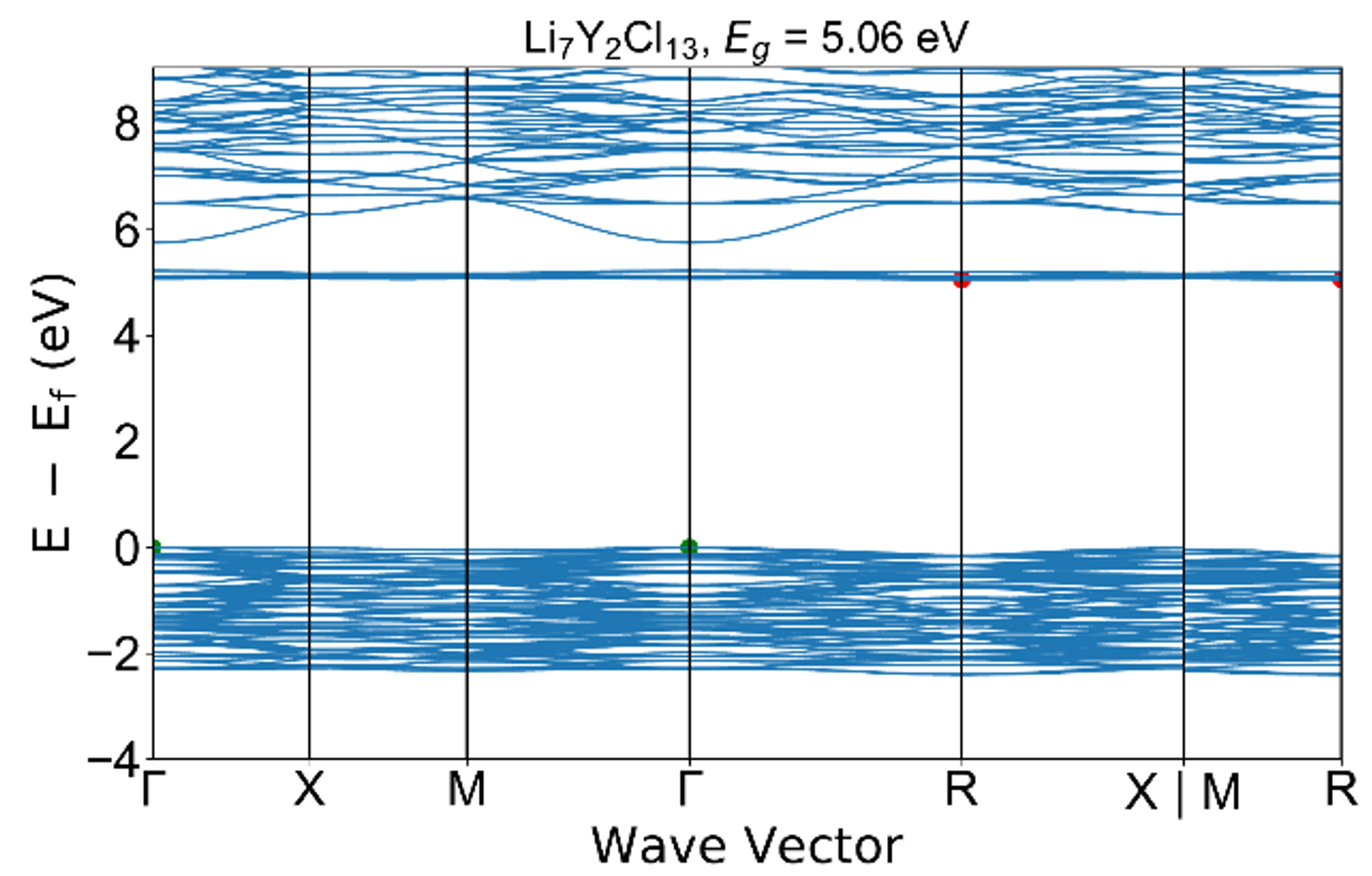}
    d.\ \includegraphics[width=0.45\textwidth]{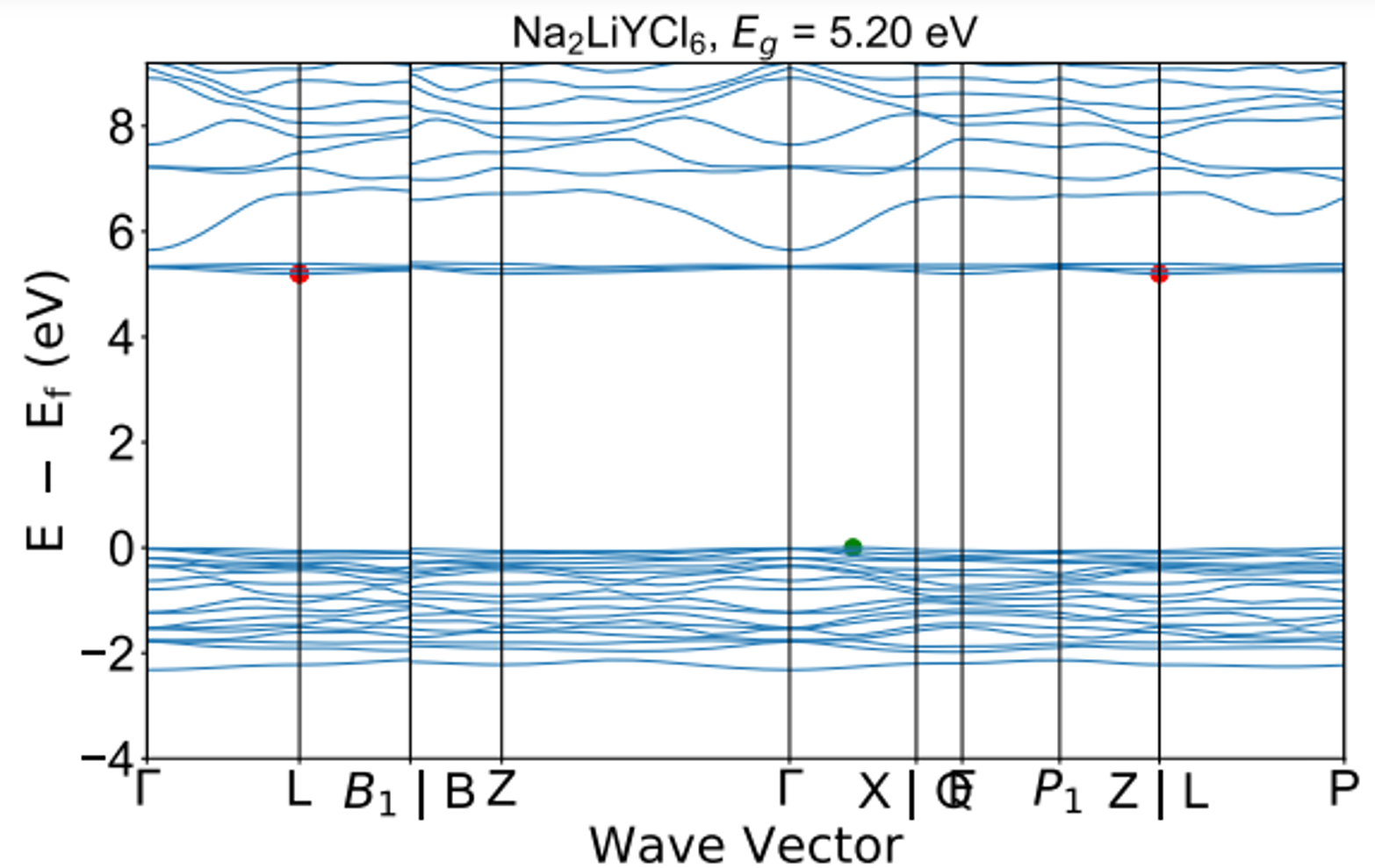}
    \caption{The \ac{DFT} \ac{PBE} band structure of the top four candidates.} \label{fig:top-band-gaps}
\end{figure}

\begin{figure}
    \centering
    ~\includegraphics[width=0.9\textwidth]{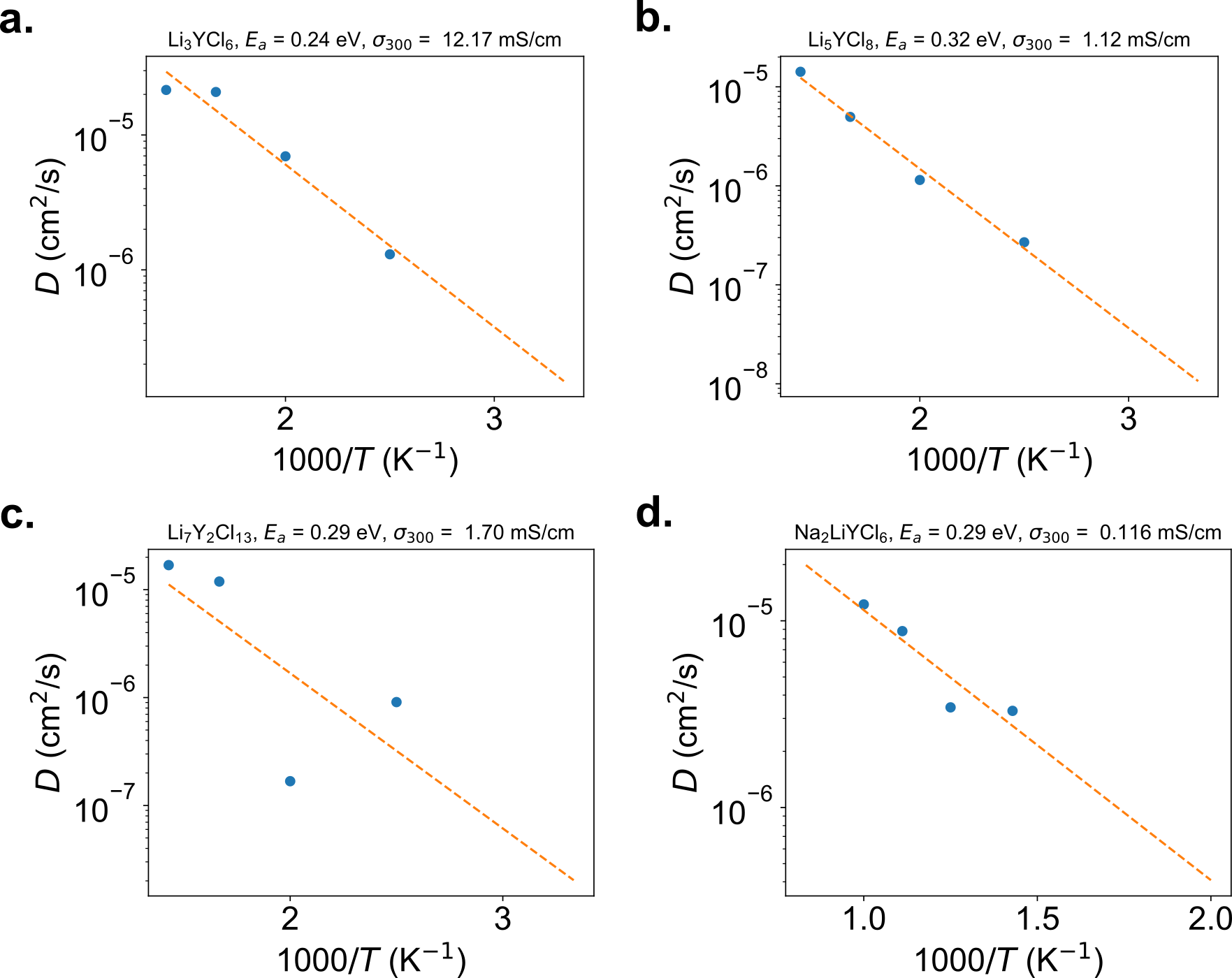}
    \caption{\Ac{AIMD} Arrhenius Li diffusivity plots for the top four candidates.} \label{fig:top-diffusivities}
\end{figure}

\begin{figure}
    \centering
    \includegraphics[width=0.55\textwidth]{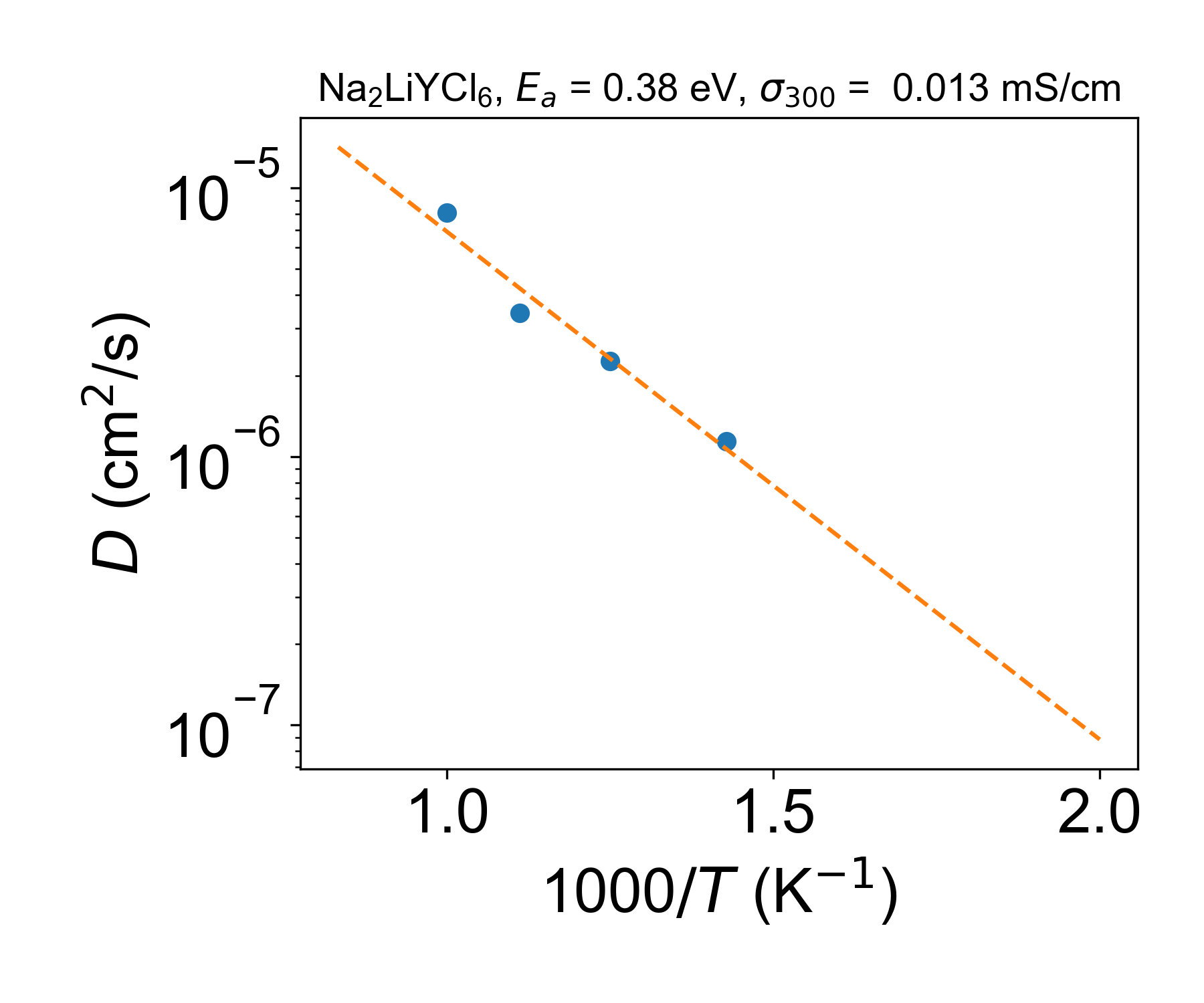}
    \caption{\Ac{AIMD} Arrhenius Na diffusivity plot for the Na$_2$LiYCl$_6$ compound.} \label{fig:n2116-na-diffusitivity}
\end{figure}

Electrochemical stability windows for the top candidates are shown in Figure~\ref{fig:top-esw}.
Band gaps for the top candidates are shown in Figure~\ref{fig:top-band-gaps}.
The Li diffusivity results from AIMD are shown in Figure~\ref{fig:top-diffusivities}.
Figure~\ref{fig:n2116-na-diffusitivity} shows Na conductivity results for the Na$_2$LiYCl$_6$ compound that contains both Li and Na.

\begin{figure}
    \centering
    \includegraphics[width=0.9\textwidth]{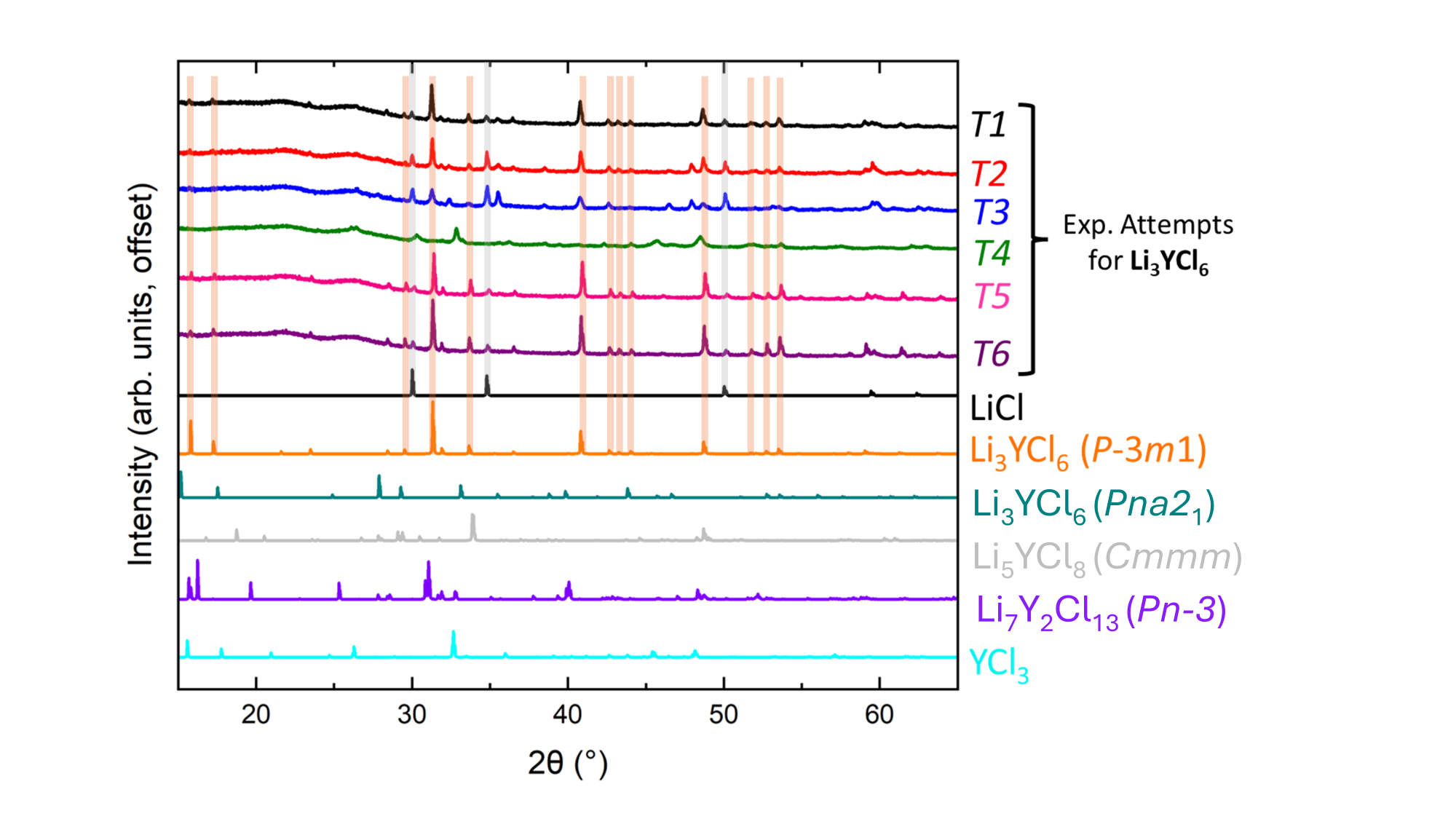}
    \caption{\Ac{PXRD} results for Li$_3$YCl$_6$ of representative reactions conducted at temperatures ranging from 250 – 650~°C (shown as T1 – T6).
    The main phases observed in these experimental screenings are Li$_3$YCl$_6$ (trigonal phase) and unreacted LiCl.} \label{fig:pxrd-li3ycl6}
\end{figure}

\begin{figure}
    \centering
    \includegraphics[width=0.9\textwidth]{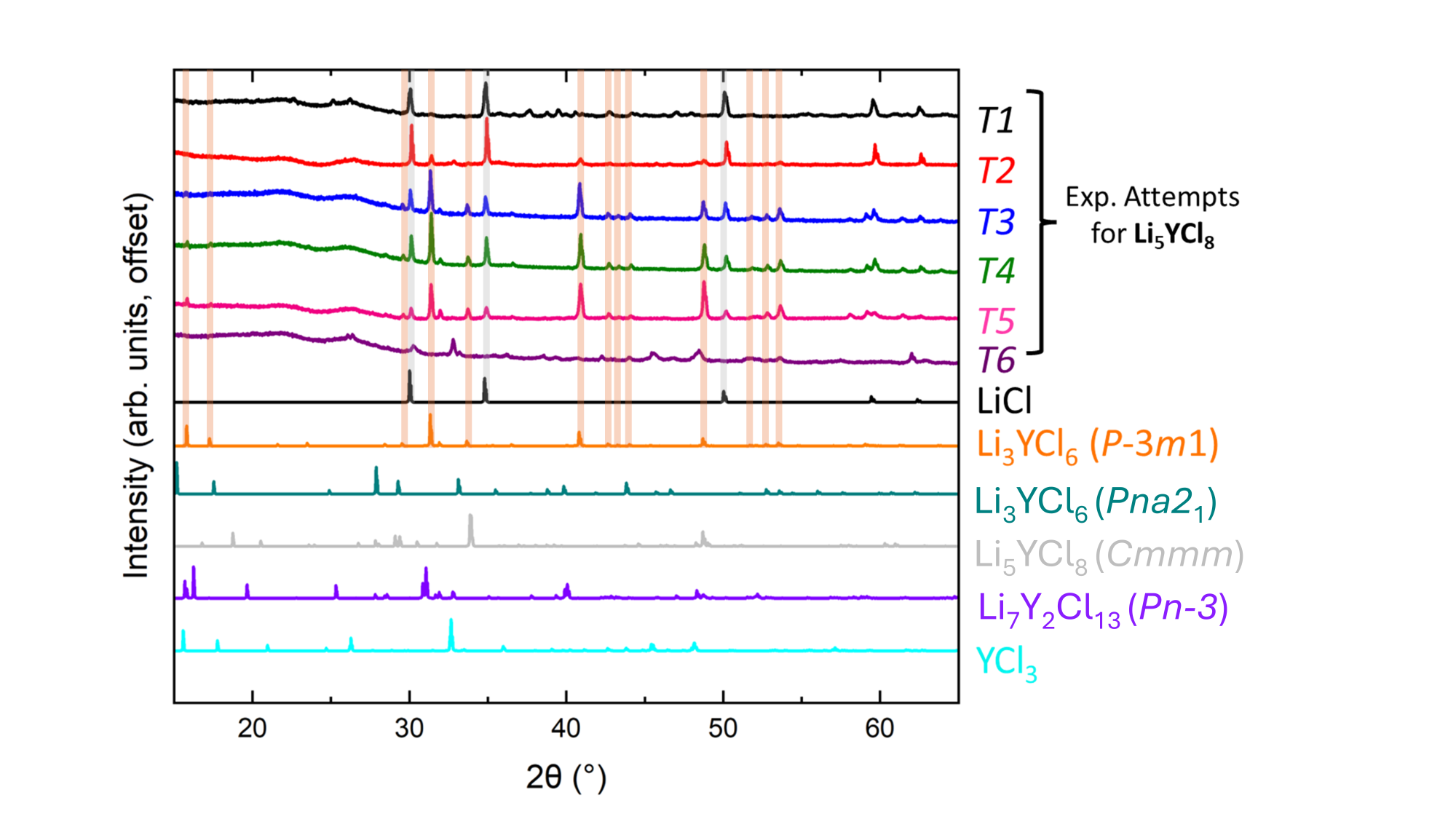}
    \caption{\Ac{PXRD} results for Li$_5$YCl$_8$ of representative reactions conducted at temperatures ranging from 250 – 650~°C (shown as T1 – T6).
    The main phases observed in these experimental screenings are Li$_3$YCl$_6$ (trigonal phase) and unreacted LiCl.} \label{fig:pxrd-li5ycl8}
\end{figure}

\begin{figure}
    \centering
    \includegraphics[width=0.9\textwidth]{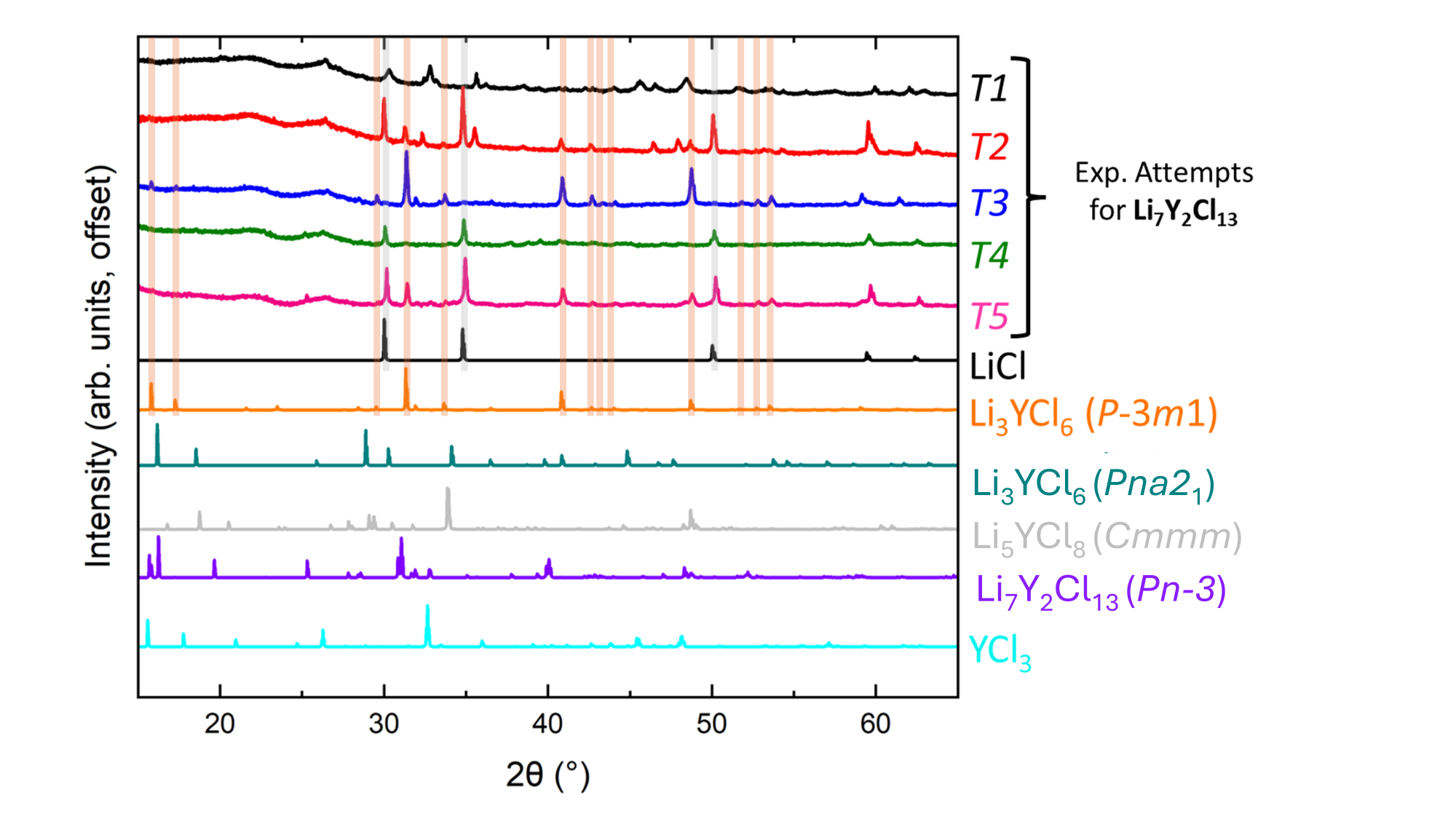}
    \caption{\Ac{PXRD} results for Li$_7$Y$_2$Cl$_{13}$ of representative reactions conducted at temperatures ranging from 250 – 650~°C (shown as T1 – T6).
    The main phases observed in these experimental screenings are Li$_3$YCl$_6$ (trigonal phase) and unreacted LiCl.} \label{fig:pxrd-li7y2cl13}
\end{figure}

\begin{figure}
    \centering
    \includegraphics[width=0.9\textwidth]{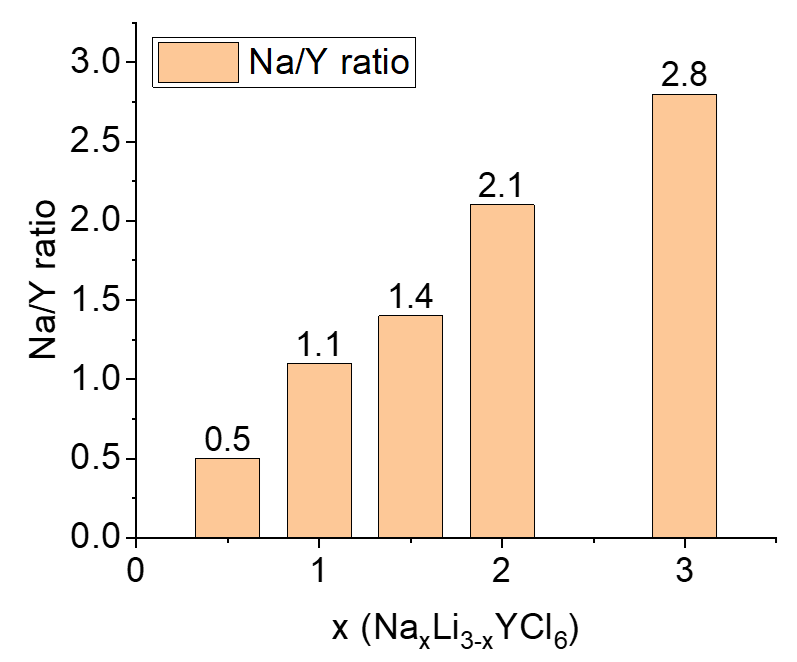}
    \caption{Na/Y ratios in \naseries\ electrolytes calculated from \ac{EDS} elemental analysis.} \label{fig:eds-ratios}
\end{figure}

Table~\ref{tab:elemental-data} shows the elemental compositions of the \naseries\ series obtained from \ac{SEM}/\ac{EDS} analysis.
\Ac{PXRD} patterns are shown in Figures~\ref{fig:pxrd-li3ycl6}, \ref{fig:pxrd-li5ycl8}, and \ref{fig:pxrd-li7y2cl13}.
Figure~\ref{fig:eds-ratios} show the elemental ratios derived from the \ac{PXRD} data.

\begin{table}
    \caption{Measured elemental compositions of the \naseries\ series obtained from \ac{SEM}/\ac{EDS} analysis, normalized to Y.
    Note that Li cannot be observed in the \ac{EDS} measurement.} \label{tab:elemental-data}
    \begin{tabular}{@{}lllll@{}}
        \toprule
        Nominal composition & Na & Li & Y & Cl \\
        \naseries & & & & \\
        \midrule
        $x = 3$   & 2.8 (2) & -- & 1 & 5.5 (3) \\
        $x = 2$   & 2.1 (2) & -- & 1 & 5.0 (5) \\
        $x = 1.5$ & 1.4 (6) & -- & 1 & 5.0 (4) \\
        $x = 1$   & 1.1 (4) & -- & 1 & 4.5 (6) \\
        $x = 0.5$ & 0.5 (1) & -- & 1 & 4.3 (9) \\
        $x = 0$   & --      & -- & 1 & 4.7 (4) \\
        \botrule
    \end{tabular}
\end{table}

\end{appendices}

\end{document}